\documentclass[12pt,nofootinbib,a4paper]{article}
\pdfoutput=1

\usepackage[english]{babel}
\usepackage[utf8x]{inputenc}
\usepackage{amsmath}
\usepackage{graphicx}

 \usepackage{lscape}

\usepackage{fancyhdr}
\usepackage{rotating}
\usepackage{amsmath}
\usepackage{mathtools}
\usepackage{amsthm}
\usepackage{amsfonts}
\usepackage[dvips]{epsfig}
\usepackage{graphicx}
\usepackage{psfrag}

\usepackage{amssymb}
\usepackage{cancel}
\usepackage{pstricks} 
\usepackage{lscape}
\usepackage{cancel}
\usepackage{slashed}
\usepackage{pstricks} 
\usepackage{lscape}
\usepackage{color}
\usepackage{cite}
\usepackage{hyperref}
\usepackage{url}
\usepackage{caption}
\usepackage{bbm}
\usepackage{bm}
\usepackage{subfig}
\usepackage{bbold}
\usepackage{caption}

\newcommand{\beq}{\begin{equation}}
\newcommand{\eeq}{\end{equation}}
\newcommand{\ga}{\lower.7ex\hbox{$\;\stackrel{\textstyle>}{\sim}\;$}}
\newcommand{\la}{\lower.7ex\hbox{$\;\stackrel{\textstyle<}{\sim}\;$}}

\newcommand{\be}{\begin{equation}}
\newcommand{\ee}{\end{equation}}
\newcommand{\bea}{\begin{eqnarray}}
\newcommand{\eea}{\end{eqnarray}}

 \def\lra#1{\overset{\text{\scriptsize$\leftrightarrow$}}{#1}}

\makeatletter
\newcommand{\Cen}[2]{%
  \ifmeasuring@
    #2%
  \else
    \makebox[\ifcase\expandafter #1\maxcolumn@widths\fi]{$\displaystyle#2$}%
  \fi
}
\makeatother

\begin{document}

\begin{titlepage}
\pagestyle{empty}

\begin{flushright}
{\tt Cavendish-HEP-2018-06,  DAMTP-2018-12,} \\
{\tt KCL-PH-TH/2018-12}, {\tt CERN-PH-TH/2018-042}
\end{flushright}
%%%%%%%%%%%%%%%%%%%%%% F I G U R E %%%%%%%%%%%%%%%%%%%%%%%%%%%%%%%%%%%

\vspace{0.3cm}
\begin{center}
{\bf {\LARGE Updated Global SMEFT Fit to Higgs, \\
\vspace {0.3cm}
Diboson and Electroweak Data}}
\end{center}

\vspace{0.05in}

\begin{center}
{\bf John~Ellis}$^{a,b}$,
{\bf Christopher~W.~Murphy}$^c$,
{\bf Ver\'onica Sanz}$^{d}$ and
{\bf Tevong~You}$^{e}$
\end{center}

{\small

\begin{center}
{\em $^a$ Theoretical Particle Physics and Cosmology Group, Department of
  Physics, King's~College~London, London WC2R 2LS, United Kingdom}\\[0.2cm]
{\em $^b$ National Institute of Chemical Physics \& Biophysics, R{\" a}vala 10, 10143 Tallinn, Estonia; \\
Theoretical Physics Department, CERN, CH-1211 Geneva 23,
  Switzerland}\\[0.2cm]
{\em $^c$ Department of Physics, Brookhaven National Laboratory, Upton, New York, 11973, USA}\\[0.2cm]
{\em $^d$ Department of Physics and Astronomy, University of Sussex, Brighton BN1 9QH, UK}\\[0.2cm]
{\em $^e$ DAMTP, University of Cambridge, Wilberforce Road, Cambridge, CB3 0WA, UK;
Cavendish Laboratory, University of Cambridge, J.~J.~Thomson Avenue, Cambridge, CB3~0HE,
 UK}
\end{center}

}

\bigskip
\bigskip

\centerline{\bf Abstract}
\vspace{0.5cm}
%\noindent  
{The ATLAS and CMS collaborations have recently released significant new data on Higgs and diboson production in LHC Run 2. Measurements of Higgs properties have improved in many channels, while kinematic information for $h \to \gamma\gamma$ and $h \to ZZ$ can now be more accurately incorporated in fits using the STXS method, 
and $W^+ W^-$ diboson production at high $p_T$ gives new sensitivity to deviations from the Standard Model. 
We have performed an updated global fit to precision electroweak data, $W^+W^-$ measurements at LEP,
and Higgs and diboson data from Runs 1 and 2 of the LHC in the framework of the
Standard Model Effective Field Theory (SMEFT), allowing all coefficients to vary %JE
across the combined dataset, and present the results in both the Warsaw and SILH operator bases.
We exhibit the improvement in the constraints on operator coefficients provided by the LHC Run 2 data, and discuss the correlations
between them. We also explore the constraints our fit results impose on several models of physics beyond the Standard Model,
including models that contribute to the operator coefficients at the tree level and stops in the MSSM that contribute via loops.}

\vspace{0.17in}

\begin{flushleft}
{\small 
March 2018
}
\end{flushleft}
\medskip
\noindent

\end{titlepage}

\newpage

\section{Introduction}

In the absence (so far) of any clear signature of some physics beyond the Standard Model (BSM) at the LHC,
the Standard Model Effective Field Theory (SMEFT) has emerged as one of the most
interesting tools to probe systematically the data from the LHC and elsewhere for hints of possible BSM physics\footnote{See Refs.~\cite{deFlorian:2016spz, Brivio:2017vri, Falkowski:2015fla, Willenbrock:2014bja} for some recent reviews of the SMEFT.}.
The formulation of the SMEFT assumes that all the known particles have the same SU(3)$_c \times$SU(2)$_L \times$U(1)$_Y$
gauge transformation properties as in the Standard Model Model (SM), with their conventional dimension-2 and -4 interactions being
supplemented by higher-dimensional interactions between all allowed combinations of these SM fields.
Such interactions might be generated by massive particles being exchanged at the tree-level
or circulating in loop diagrams. These interactions would in general be suppressed by powers of some
high mass scale $\Lambda$ related to the scale of BSM  physics, with dimensionless coefficients that depend
on their interactions with SM particles. The leading higher-dimensional operators relevant to many LHC
measurements are expected to be those of dimension 6. If the LHC experiments measure one or more significant
deviations from SM predictions, the SMEFT can be used to help characterize its possible origin. In the absence of any
significant deviations, the SMEFT can be used to constrain the scales of different BSM physics scenarios and to guide the search for direct effects of new physics. %JE

First steps in this SMEFT programme have included the cataloguing of possible interactions of dimension 5, 6
and higher~\cite{Weinberg:1979sa, Buchmuller:1985jz, Lehman:2014jma, Henning:2015alf}, the construction of non-redundant bases of independent operators~\cite{Grzadkowski:2010es, Contino:2013kra, Gupta:2014rxa, Masso:2014xra, deFlorian:2016spz}, and the development of a dictionary to translate between different bases~\cite{Falkowski:2015wza, Aebischer:2017ugx}. This groundwork has been the basis for subsequent phenomenological
analyses through global fits of data~\cite{Han:2004az, Pomarol:2013zra, Ellis:2014dva, Ellis:2014jta, Murphy:2017omb, Falkowski:2014tna, Efrati:2015eaa, Falkowski:2015jaa, Falkowski:2015krw, Falkowski:2016cxu, Falkowski:2017pss, Falkowski:2018dmy, Corbett:2012ja, Corbett:2013pja, Englert:2015hrx, Buckley:2016cfg, Buckley:2015lku, Corbett:2015ksa, Butter:2016cvz, Dumont:2013wma, deBlas:2017wmn, deBlas:2016nqo, Berthier:2015oma, Berthier:2015gja, Berthier:2016tkq, Brivio:2017bnu} from the LHC and other experiments that constrain various combinations of dimension-6
operator coefficients and thereby different scenarios for BSM physics. The principal classes of observables
used in such analyses have included precision electroweak data from LEP~\cite{ALEPH:2005ab}, the SLC~\cite{ALEPH:2005ab} and the Tevatron~\cite{Aaltonen:2013iut
},
constraints on diboson production from LEP~2~\cite{Heister:2004wr, Achard:2004zw, Abbiendi:2007rs, Schael:2013ita} and the LHC~\cite{Aaboud:2017qkn, Dorigo:2018cbl}, and data on Higgs production from
the LHC~\cite{Khachatryan:2016vau}. In the past, the precision of the electroweak $Z$-pole data has been such that the coefficients of
the operators affecting them could initially be considered independently of those entering into other observables.
However, such a segregated approach is theoretically unsatisfactory, with some bases being more correlated across measurements than others, and is becoming obsolescent with the advent
of more precise LHC data on Higgs production and diboson production where the latter, in particular, %JE
can no longer be interpreted solely as a measurement of anomalous triple-gauge couplings~\cite{Zhang:2016zsp, Baglio:2017bfe}.

In this paper we perform the first comprehensive global analysis of relevant electroweak and diboson data together with Higgs production data from Runs 1 and 2 of the LHC, while allowing all relevant operators to vary %JE
in the combined dataset, thus superseding our previous analyses~\cite{Ellis:2014dva, Ellis:2014jta, Murphy:2017omb}. As we discuss in more detail below, we
include in our analysis 14 precision electroweak measurements, 74 measurements of $e^+ e^- \to W^+ W^- \to 4$
fermions, 22 Higgs signal strength measurements from Run~1 of the LHC and 46 measurements of Higgs
production from Run~2 of the LHC (including information using Simplified Template Cross Sections (STXS)~\cite{deFlorian:2016spz}), and one measurement of $W^+W^-$ production at high $p_T$ during Run~2 of the LHC.

We present our results in both the Warsaw~\cite{Grzadkowski:2010es} and SILH~\cite{Giudice:2007fh, Contino:2013kra} operator bases and in two forms:
one in which all the dimension-6 operator coefficients are allowed to be non-vanishing simultaneously, and one in which
the operator coefficients are switched on one at a time. We exhibit the improvement in the constraints on operator
coefficients compared to fits using only data from Run 1 of the LHC, and we discuss the correlations between the constraints
on the coefficients. We also analyze the implications of this fit for BSM models that make tree-level
contributions to the operator coefficients~\cite{deBlas:2017xtg}, %JE
as well as for stop squarks in the minimal supersymmetric extension of the SM
(MSSM), which contribute to the operator coefficients at the loop level~\cite{Henning:2014wua,Henning:2014gca,Drozd:2015kva}. %JE

The layout of our paper is as follows. In section~2 we review the SMEFT framework and introduce the 20
dimension-6 operators that appear in our analysis, and in Section~3 we introduce the data we use.
Section~4 presents the methodology we use for our fit, and Section~5 presents the results of our analysis.
Their implications for a variety of BSM scenarios are discussed in Section~6. Finally, Section~7 summarizes our conclusions.

\section{The SMEFT Framework}

The SM is defined by a Lagrangian consisting of all operators up to mass dimension 4 formed by combinations of SM fields that are allowed by a linearly-realized SU(3)$_c \times$SU(2)$_L \times$U(1)$_Y$ gauge symmetry. However, if new physics exists at some heavier scale $\Lambda$, one generically expects higher-dimensional operators to also be present, their effects suppressed by $\Lambda$ to powers fixed by dimensional analysis, with logarithmic corrections that are calculable in perturbation theory. Treating the SM properly as a low-energy Effective Field Theory (EFT), the SMEFT is the SM Lagrangian extended to include a series of higher-dimensional operators. At dimension 5 there is a single category of operators, which violate lepton number and give masses to neutrinos~\cite{Weinberg:1979sa}. %JE
Here we focus on the effects of the leading lepton-number-conserving operators $\mathcal{O}$ of dimension 6, 
\begin{equation}
\mathcal{L}_\text{SMEFT} \supset \mathcal{L}_\text{SM} + \sum_i \frac{c_i}{\Lambda_i^2}\mathcal{O}_i \, ,
\end{equation}
where the $c_i$ are Wilson coefficients induced by integrating out the heavy degrees of freedom of some new physics at a scale $\Lambda$\footnote{For some recent developments on matching using functional methods, see Refs.~\cite{Henning:2014wua, Drozd:2015rsp, delAguila:2016zcb, Boggia:2016asg, Henning:2016lyp, Ellis:2016enq, Fuentes-Martin:2016uol, Zhang:2016pja, Ellis:2017jns}.}. One would typically expect a tree-level contribution to be proportional to at least the square
of some new physics coupling, e.g.,  $c \sim g_*^2$, with an additional suppression by a %JE
factor $\sim 1 / (4\pi)^2$    if it appears when the BSM physics is integrated out at one loop. From a bottom-up point of view the coefficients are treated as free parameters where the validity of the EFT can be assessed {\it a posteriori}~\cite{Contino:2016jqw}.

The coefficients $c(\Lambda)$ generated at the scale $\Lambda$ are related to their values $c(v)$ at the electroweak scale $v \sim 246$ GeV through RGE running, using the SMEFT one-loop anomalous dimension matrix that has been calculated in Refs.~\cite{Grojean:2013kd, Elias-Miro:2013gya, Elias-Miro:2013mua, Jenkins:2013zja, Jenkins:2013wua, Alonso:2013hga, Alonso:2014zka}. Below the electroweak scale the SMEFT can be matched to a %JE
low-energy EFT~\cite{Jenkins:2017jig, Celis:2017hod, Aebischer:2015fzz, Aebischer:2017gaw}, whose running is also known~\cite{Jenkins:2017dyc}. Since the data currently
do not require a large hierarchy between the electroweak scale and the BSM scale, and we work to leading order for simplicity\footnote{See Refs.~\cite{Brivio:2017vri, Trott:2017yhn, Hartmann:2016pil, Passarino:2016pzb, Hartmann:2015aia, Gauld:2016kuu, Dawson:2018pyl} for some discussion and results in the SMEFT at NLO.}, we do not discuss these effects in this paper. 

The dimension-6 operators were first classified systematically in Ref.~\cite{Buchmuller:1985jz}. Such a list generally forms a redundant set since operators related by field redefinitions, equations of motion, integration by parts, or Fierz identities give identical $S$-matrix predictions and are therefore equivalent descriptions of the same physics\footnote{The problem of generating a non-redundant set of operators to arbitrary mass dimension has recently been solved using Hilbert series methods~\cite{Lehman:2015via, Lehman:2015coa, Henning:2015daa, Henning:2015alf, Henning:2017fpj}.}. The first non-redundant basis of operators was derived in Ref.~\cite{Grzadkowski:2010es} and is commonly called the Warsaw basis. Another popular basis in the literature is referred to as the SILH basis~\cite{Giudice:2007fh, Contino:2013kra}. There are 2499 CP-even dimension-6 operators, which reduce to 59 independent operators when assuming minimal flavour violation~\cite{Alonso:2013hga}, but of those only 20 are relevant for the Higgs, diboson, and electroweak precision observables that we consider here\footnote{We do not consider CP-odd operators in our analysis; for a recent study of CP tests in the Higgs sector see Ref.~\cite{Brehmer:2017lrt}.}. 
We assume here an U(3)$^5$ flavor symmetry, under which the Yukawa matrices are promoted to spurions transforming as bi-triplets, and present results in both the Warsaw and SILH bases. %JE

In the Warsaw basis, the 11 operators involved in diboson measurements and electroweak precision observables, through input parameter shifts or modifications of the gauge boson self-coupling and couplings to fermions, can be written in the notation of Ref.~\cite{Brivio:2017vri} as  
{\small
\begin{align}
\mathcal{L}_\text{SMEFT}^\text{Warsaw} &\supset \frac{\bar{C}_{Hl}^{(3)}}{v^2} (H^\dag i\overleftrightarrow{D}^I_\mu H)(\bar l \tau^I \gamma^\mu l) + \frac{\bar{C}_{Hl}^{(1)}}{v^2}(H^\dag i\overleftrightarrow{D}_\mu H)(\bar l \gamma^\mu l) +\frac{\bar{C}_{ll}}{v^2}(\bar l \gamma_\mu l)(\bar l \gamma^\mu l) \nonumber \\
&+ \frac{\bar{C}_{HD}}{v^2}\left|H^\dag D_\mu H\right|^2 + \frac{\bar{C}_{HWB}}{v^2} H^\dag \tau^I H\, W^I_{\mu\nu} B^{\mu\nu}  \nonumber  \\ 
&+ \frac{\bar{C}_{He}}{v^2} (H^\dag i\overleftrightarrow{D}_\mu H)(\bar e \gamma^\mu e) \nonumber + \frac{\bar{C}_{Hu}}{v^2} (H^\dag i\overleftrightarrow{D}_\mu H)(\bar u \gamma^\mu u) + \frac{\bar{C}_{Hd}}{v^2} (H^\dag i\overleftrightarrow{D}_\mu H)(\bar d \gamma^\mu d) \nonumber \\
&+ \frac{\bar{C}_{Hq}^{(3)}}{v^2} (H^\dag i\overleftrightarrow{D}^I_\mu H)(\bar q \tau^I \gamma^\mu q) + \frac{\bar{C}_{Hq}^{(1)}}{v^2} (H^\dag i\overleftrightarrow{D}_\mu H)(\bar q \gamma^\mu q) + \frac{\bar{C}_{W}}{v^2} \epsilon^{IJK} W_\mu^{I\nu} W_\nu^{J\rho} W_\rho^{K\mu} \, ,
%+ \frac{\bar{C}_H}{v^2} (H^\dag H)^3
\end{align}
}
where flavour indices and Hermitian conjugate operators are implicit,\footnote{ The flavour indices of the four-lepton operator are $C_{ll} = C_{\substack{ll \\ e\mu\mu e}} = C_{\substack{ll \\ \mu ee\mu}}$.} and we defined
\begin{equation}
\bar{C} \equiv \frac{v^2}{\Lambda^2}C \, .
\end{equation}
There are in addition 9 operators that affect Higgs measurements, 
{\small
\begin{align}
\mathcal{L}_\text{SMEFT}^\text{Warsaw} &\supset  \frac{\bar{C}_{eH}}{v^2} y_e (H^\dag H)(\bar l e H) + \frac{\bar{C}_{dH}}{v^2} y_d (H^\dag H)(\bar q d H) + \frac{\bar{C}_{uH}}{v^2} y_u (H^\dag H)(\bar q u \widetilde H) \nonumber \\
&+ \frac{\bar{C}_{G}}{v^2} f^{ABC} G_\mu^{A\nu} G_\nu^{B\rho} G_\rho^{C\mu}  + \frac{\bar{C}_{H\Box}}{v^2} (H^\dag H)\Box(H^\dag H) + \frac{\bar{C}_{uG}}{v^2} y_u (\bar q \sigma^{\mu\nu} T^A u) \widetilde H \, G_{\mu\nu}^A   \nonumber \\
&+ \frac{\bar{C}_{HW}}{v^2} H^\dag H\, W^I_{\mu\nu} W^{I\mu\nu} + \frac{\bar{C}_{HB}}{v^2}H^\dag H\, B_{\mu\nu} B^{\mu\nu} + \frac{\bar{C}_{HG}}{v^2} H^\dag H\, G^A_{\mu\nu} G^{A\mu\nu} \, .
\end{align}}
The $\mathcal{O}_H =|H|^6$ operator, not listed here, can be measured in double-Higgs production, for which there is %JE
limited sensitivity at the LHC\footnote{Prospects for future double-Higgs measurements at higher luminosity or energy are studied, for example, in Refs.~\cite{Kim:2018uty, Azatov:2015oxa, Baglio:2012np, DiVita:2017eyz, Alves:2017ued, Dawson:2017vgm, DiVita:2017vrr} and references therein.}. %JE

We note that Higgs production in association with a top-quark pair probes many coefficients in the SMEFT~\cite{AguilarSaavedra:2018nen, Buckley:2016cfg, Buckley:2015lku} but a number of these do not appear in our other observables---the only one we consider explicitly here is $C_{uG}$,
which is expected to make the largest contribution to $t \bar{t} h$ production. However, it should %JE
be borne in mind that the bounds on $C_{uG}$ in this work are actually bounds on the following linear combination of coefficients,
\begin{equation}
C_{uG} \rightarrow C_{uG} + 0.006 C_{uW} + 0.002 C_{uB} - 0.13 C_{qu}^{(8)} + \text{ additional } \psi^4 \text{ operators} \, .
\end{equation}
We note also that Higgs production in association with a jet is sensitive to the triple-gluon operator.
Although we will sometimes include $C_G$, or equivalently $\bar{c}_{3G}$ for the SILH basis, in our fits, more stringent bounds on $C_G$ have been derived from multi-jet processes at the LHC~\cite{Krauss:2016ely}. Other SMEFT operators that do not appear here can be constrained independently of Higgs, diboson, and electroweak precision measurements. %JE

In the SILH basis the relevant operators for our fit, with conventions defined in~\cite{Contino:2013kra} (which differs slightly from Ref.~\cite{Alloul:2013naa}), are 
\begin{align}
\mathcal{L}_\text{SMEFT}^\text{SILH} &\supset \frac{\bar{c}_W}{m_W^2} \frac{ig}{2}\left( H^\dagger  \sigma^a \lra {D^\mu} H \right )D^\nu  W_{\mu \nu}^a + \frac{\bar{c}_B}{m_W^2}\frac{ig'}{2}\left( H^\dagger  \lra {D^\mu} H \right )\partial^\nu  B_{\mu \nu} + \frac{\bar{c}_T}{v^2}\frac{1}{2}\left (H^\dagger {\lra{D}_\mu} H\right)^2 \nonumber \\
& + \frac{\bar{c}_{ll}}{v^2} (\bar L \gamma_\mu L)(\bar L \gamma^\mu L) + \frac{\bar{c}_{He}}{v^2} (i H^\dagger {\lra { D_\mu}} H)( \bar e_R\gamma^\mu e_R) + \frac{\bar{c}_{Hu}}{v^2} (i H^\dagger {\lra { D_\mu}} H)( \bar u_R\gamma^\mu u_R) \nonumber \\
& + \frac{\bar{c}_{Hd}}{v^2} (i H^\dagger {\lra { D_\mu}} H)( \bar d_R\gamma^\mu d_R) + \frac{\bar{c}_{Hq}^\prime}{v^2} (i H^\dagger \sigma^a {\lra { D_\mu}} H)( \bar Q_L\sigma^a\gamma^\mu Q_L)  \nonumber \\
&+ \frac{\bar{c}_{Hq}}{v^2} (i H^\dagger {\lra { D_\mu}} H)( \bar Q_L\gamma^\mu Q_L) + \frac{\bar{c}_{HW}}{m_W^2} i g(D^\mu H)^\dagger\sigma^a(D^\nu H)W^a_{\mu\nu} + \frac{\bar{c}_{HB}}{m_W^2} i g^\prime(D^\mu H)^\dagger(D^\nu H)B_{\mu\nu} \nonumber \\
&+ \frac{\bar{c}_{3W}}{m_W^2} g^3 \epsilon_{abc} W^{a\, \nu}_{\mu}W^{b}_{\nu\rho}W^{c\, \rho\mu} + \frac{\bar{c}_g}{m_W^2}g_s^2 |H|^2 G_{\mu\nu}^A G^{A\mu\nu} + \frac{\bar{c}_\gamma}{m_W^2}{g}^{\prime 2} |H|^2 B_{\mu\nu}B^{\mu\nu} \nonumber \\
& + \frac{\bar{c}_H}{v^2}\frac{1}{2}(\partial^\mu |H|^2)^2 + \sum_{f=e,u,d} \frac{\bar{c}_f}{v^2} y_f |H|^2    \bar{F}_L H^{(c)} f_R \nonumber \\
& + \frac{\bar{c}_{3G}}{m_W^2}g_s^3 f_{ABC} G_\mu^{A\nu} G_\nu^{B\rho} G_\rho^{C\mu} + \frac{\bar{c}_{uG}}{m_W^2} g_s y_u\bar{Q}_L H^{(c)} \sigma^{\mu\nu}\lambda_A u_R G^A_{\mu\nu}	\, .
\end{align}
Hermitian conjugates and flavour indices are again kept implicit. 

Our computations are performed at linear order in 
the Warsaw and SILH bases using $\alpha$, $G_F$, and $M_Z$ as input parameters. %JE
We used the predictions for electroweak precision observables and $WW$ scattering at LEP 2 in the Warsaw basis from Refs.~\cite{Berthier:2016tkq, Brivio:2017vri}.
Predictions for LHC observables are made using $\mathtt{SMEFTsim}$~\cite{Brivio:2017btx}.
These computations can be converted to the SILH basis using the known results in the literature~\cite{Alonso:2013hga, Henning:2014wua, Falkowski:2015wza}.

\section{Data used in the Global Fit}
\label{sec:data}

The following data are used in our global fit, which, as stated above, are sensitive to 20 directions in the SMEFT parameter space.

\begin{itemize}
\item \textit{Precision Electroweak Data}:  
We use the $Z$-pole observables from Table 8.5 of Ref.~\cite{ALEPH:2005ab}, including the correlations.
We use the $W$ mass measurements from the Tevatron~\cite{Aaltonen:2013iut} and ATLAS~\cite{Aaboud:2017svj}.
These measurements and the corresponding theoretical predictions within the SM are summarized in Table~\ref{tab:ewpd}, and they probe eight directions in the SMEFT.

\item $e^+ e^- \to W^+ W^- \to 4\, fermions$: 
We use all the data from Tables 12, 13, 14, and 15 of Ref.~\cite{Berthier:2016tkq}. 
The original experimental results can be found in Refs.~\cite{Heister:2004wr, Achard:2004zw, Abbiendi:2007rs, Schael:2013ita}, and
we use the SM predictions from Ref.~\cite{Achard:2004zw, Abbiendi:2007rs}. 
This is a total of 74 measurements.
These measurements also probe eight directions in the SMEFT.
However only three of these combinations of the parameters are unconstrained by the electroweak precision data.

\item \textit{Higgs Production in LHC Run 1}: 
We use all the 20 signal strengths from Table 8 of Ref.~\cite{Khachatryan:2016vau}, including the correlations given in Figure 27 of the same paper, where a signal strength is defined as the ratio of the measured cross section to its SM prediction. 
The ATLAS and CMS combination for the $h \to \mu^+ \mu^-$ signal strength is taken from Table 13 of Ref.~\cite{Khachatryan:2016vau}. 
The ATLAS $h \to Z \gamma$ signal strength is taken from Figure 1 of Ref.~\cite{Aad:2015gba}. 
These measurements are summarized in Table~\ref{tab:Hrun1}.
The 20 correlated measurements are sensitive to nine combinations of SMEFT parameters,
and the measurement of $h \to Z \gamma$ constitutes a tenth direction.
However, $h \to \mu^+ \mu^-$ is a dependent quantity because of the $U(3)^5$ flavor symmetry that we assume. %JE

\item \textit{Higgs Production in LHC Run 2}: 
We use 25 measurements from CMS~\cite{Sirunyan:2017dgc, Sirunyan:2017elk, Sirunyan:2018mvw, Sirunyan:2018shy, Sirunyan:2018shy, CMS-PAS-HIG-16-042, Sirunyan:2018ouh, Sirunyan:2017exp, Sirunyan:2017khh}, and 23 measurements from ATLAS~\cite{Aaboud:2017ojs, Aaboud:2017xsd, Aaboud:2017rss, Aaboud:2017jvq, ATLAS-CONF-2018-004, ATLAS-CONF-2017-047, ATLAS-CONF-2016-112}. 
A summary is given in Table~\ref{tab:Hrun2}.
The correlations between the $4\ell$ and $\gamma\gamma$ decay notes from Ref.~\cite{ATLAS-CONF-2017-047} are also included in the context of template cross sections (STXS) as described in Ref.~\cite{Hays:2290628}. 
These measurements probe 12 combinations of SMEFT parameters\footnote{A SMEFT fit to ATLAS
Higgs production data is presented in~\cite{ATLAS-CONF-2017-047}. See also a recent non-linear EFT analysis in~\cite{deBlas:2018tjm}
and a global SM fit to electroweak and Higgs measurements in~\cite{Haller:2018nnx}.}. %JE

\item $W^+ W^-$ \textit{Production at the LHC}: 
We use only one measurement of the differential cross section for $p p \to W W \to e^{\pm} \nu \mu^{\mp} \nu$ by ATLAS at 13~TeV~\cite{Aaboud:2017qkn} as no correlations are provided. 
The particular bin we chose, which requires the transverse momentum ($p_T$) of the leading lepton ($\ell1$) to be greater than 120~GeV, is the overflow bin, which is expected to maximize the sensitivity to certain Wilson coefficients. 
The signal strength for this measurement is $\mu(p p \to e^{\pm} \nu \mu^{\mp} \nu; p_T^{\ell 1} > 120~\text{GeV}) = 1.05 \pm 0.06 (\text{exp.}) \pm 0.1 (\text{theo.})$.
\end{itemize}

%%%%%%
\begin{table}
\begin{center}
 \begin{tabular}{| c | c | c | c | c |}
 \hline 
Observable & Measurement & Ref. & SM Prediction & Ref.  \\ \hline \hline
$\Gamma_Z$~[GeV] & $2.4952 \pm 0.0023$ & \cite{ALEPH:2005ab} & $2.4943 \pm 0.0005$ & \cite{Brivio:2017bnu} \\ \hline
$\sigma_{\text{had}}^0$~[nb] & $41.540 \pm 0.037$ & \cite{ALEPH:2005ab} & $41.488 \pm 0.006$ &  \cite{Brivio:2017bnu} \\ \hline
$R_{\ell}^0$ & $20.767 \pm 0.025$ & \cite{ALEPH:2005ab} & $20.752 \pm 0.005$ &  \cite{Brivio:2017bnu} \\ \hline
$A_{\text{FB}}^{0, \ell}$ & $0.0171 \pm 0.0010 $ & \cite{ALEPH:2005ab} & $0.01622 \pm 0.00009$ &  \cite{Patrignani:2016xqp} \\ \hline
$\mathcal{A}_{\ell}\left(P_{\tau}\right)$ & $0.1465 \pm 0.0033 $ & \cite{ALEPH:2005ab} & $0.1470 \pm 0.0004$ &  \cite{Patrignani:2016xqp} \\ \hline
$\mathcal{A}_{\ell}\left(\text{SLD}\right)$ & $0.1513 \pm 0.0021 $ & \cite{ALEPH:2005ab} & $0.1470 \pm 0.0004$ &  \cite{Patrignani:2016xqp} \\ \hline
$R_{b}^0$ & $0.021629 \pm 0.00066$ & \cite{ALEPH:2005ab} & $0.2158 \pm 0.00015$ & \cite{Brivio:2017bnu} \\ \hline
$R_{c}^0$ & $0.1721 \pm 0.0030$ & \cite{ALEPH:2005ab} & $0.17223 \pm 0.00005$ & \cite{Brivio:2017bnu} \\ \hline
$A_{\text{FB}}^{0, b}$ & $0.0992 \pm 0.0016$ & \cite{ALEPH:2005ab} & $0.1031 \pm 0.0003$ & \cite{Patrignani:2016xqp} \\ \hline
$A_{\text{FB}}^{0, c}$ & $0.0707 \pm 0.0035$ & \cite{ALEPH:2005ab} & $0.0736 \pm 0.0002$ & \cite{Patrignani:2016xqp} \\ \hline
$\mathcal{A}_b$ & $0.923 \pm 0.020$ & \cite{ALEPH:2005ab} & $0.9347$ & \cite{Patrignani:2016xqp} \\ \hline
$\mathcal{A}_c$ & $0.670 \pm 0.027$ & \cite{ALEPH:2005ab} & $0.6678 \pm 0.0002$ & \cite{Patrignani:2016xqp} \\ \hline
$M_W$~[GeV] & $80.387 \pm 0.016$ & \cite{Aaltonen:2013iut} & $80.361 \pm 0.006$ & \cite{Patrignani:2016xqp} \\ \hline
$M_W$~[GeV] & $80.370 \pm 0.019$ & \cite{Aaboud:2017svj} & $80.361 \pm 0.006$ & \cite{Patrignani:2016xqp} \\ \hline
 \end{tabular}
 \end{center}
  \caption{\it Summary of the precision electroweak data used in our global fit.}
  \label{tab:ewpd}
\end{table}

%%%%%%
\begin{table}
\begin{center}
 \begin{tabular}{| c | c | c || c | c | c |}
 \hline 
Production & Decay & Signal Strength & Production & Decay & Signal Strength  \\ \hline \hline
$gg$F & $\gamma\gamma$ & $1.10^{+0.23}_{-0.22}$ & $Wh$ & $\tau\tau$ & $-1.4\pm1.4$ \\ \hline
$gg$F & $ZZ$ & $1.13^{+0.34}_{-0.31}$ & $Wh$ & $bb$ & $1.0\pm0.5$ \\ \hline
$gg$F & $WW$ & $0.84\pm0.17$ & $Zh$ & $\gamma\gamma$ & $0.5^{+3.0}_{-2.5}$ \\ \hline
$gg$F & $\tau\tau$ & $1.0\pm0.6$ & $Zh$ & $WW$ & $5.9^{+2.6}_{-2.2}$ \\ \hline
VBF & $\gamma\gamma$ & $1.3\pm0.5$ & $Zh$ & $\tau\tau$ & $2.2^{+2.2}_{-1.8}$ \\ \hline
VBF & $ZZ$ & $0.1^{+1.1}_{-0.6}$ & $Zh$ & $bb$ & $0.4\pm0.4$ \\ \hline
VBF & $WW$ & $1.2\pm0.4$ & $tth$ & $\gamma\gamma$ & $2.2^{+1.6}_{-1.3}$ \\ \hline
VBF & $\tau\tau$ & $1.3\pm0.4$ & $tth$ & $WW$ & $5.0^{+1.8}_{-1.7}$ \\ \hline
$Wh$ & $\gamma\gamma$ & $0.5^{+1.3}_{-1.2}$ & $tth$ & $\tau\tau$ & $-1.9^{+3.7}_{-3.3}$ \\ \hline
$Wh$ & $WW$ & $1.6^{+1.2}_{-1.0}$ & $tth$ & $bb$ & $1.1\pm1.0$ \\ \hline
$pp$ & $Z\gamma$ & $2.7^{+4.6}_{-4.5}$ & $pp$ & $\mu\mu$ & $0.1\pm2.5$ \\ \hline
 \end{tabular}
 \end{center}
  \caption{\it Summary of LHC Run 1 Higgs results used in this work. All the measurements are 
  combined CMS and ATLAS results from Ref.~\cite{Khachatryan:2016vau}, except for the $Z \gamma$ result, which is 
  an ATLAS result from Ref.~\cite{Aad:2015gba}.}
  \label{tab:Hrun1}
\end{table}

%%%%%%
\begin{table}
\hspace{-1.2cm}
%\begin{center}
% \resizebox{\textwidth}{!}{%
 \begin{tabular}{| c | c | c | c || c | c | c | c |}
 \hline 
% \multicolumn{4}{|c ||}{CMS}&\multicolumn{4}{c |}{ATLAS} \\ \hline \hline
 & Production & Decay & Sig. Stren. &  & Production & Decay & Sig. Stren. \\ \hline \hline
\cite{Sirunyan:2017dgc} & 1-jet, $p_T > 450$ & $b \bar{b}$ & $2.3^{+1.8}_{-1.6}$ & \cite{Aaboud:2017ojs} & $p p$ & $\mu \mu$ & $-0.1 \pm 1.5$ \\ \hline
\cite{Sirunyan:2017elk} & $Z h$ & $b \bar{b}$ & $0.9 \pm 0.5$ & \cite{Aaboud:2017xsd} & $Z h$ & $b \bar{b}$ & $1.12^{+0.50}_{-0.45}$ \\ \hline
\cite{Sirunyan:2017elk} & $W h$ & $b \bar{b}$ & $1.7 \pm 0.7$ &  \cite{Aaboud:2017xsd}  & $W h$ & $b \bar{b}$  & $1.35^{+0.68}_{-0.59}$ \\ \hline
\cite{Sirunyan:2018mvw} & $t \bar{t} h, \geq 1\ell$ & $b \bar{b}$ & $0.72 \pm 0.45$ & \cite{Aaboud:2017rss} & $t \bar{t} h$ & $b \bar{b}$ & $0.84^{+0.64}_{-0.61}$ \\ \hline
\cite{Sirunyan:2018shy} & $t \bar{t} h$ & $1\ell + 2 \tau_h$ & $-1.52^{+1.76}_{-1.72}$ & \cite{Aaboud:2017jvq} & $t \bar{t} h$ & $2\ell os + 1 \tau_h$ & $1.7^{+2.1}_{-1.9}$ \\ \hline
\cite{Sirunyan:2018shy} & $t \bar{t} h$ & $2\ell ss + 1 \tau_h$ & $0.94^{+0.80}_{-0.67}$ &  \cite{Aaboud:2017jvq} & $t \bar{t} h$ & $1 \ell + 2 \tau_h$ & $-0.6^{+1.6}_{-1.5}$ \\ \hline
\cite{Sirunyan:2018shy} & $t \bar{t} h$ & $3\ell + 1 \tau_h$ & $1.34^{+1.42}_{-1.07}$ & \cite{Aaboud:2017jvq} & $t \bar{t} h$ & $3 \ell + 1 \tau_h$ & $1.6^{+1.8}_{-1.3}$ \\ \hline
\cite{Sirunyan:2018shy} & $t \bar{t} h$ & $2\ell ss$ & $1.61^{+0.58}_{-0.51}$ &  \cite{Aaboud:2017jvq} & $t \bar{t} h$ & $2 \ell ss + 1 \tau_h$ & $3.5^{+1.7}_{-1.3}$ \\ \hline
\cite{Sirunyan:2018shy} & $t \bar{t} h$ & $3\ell$ & $0.82^{+0.77}_{-0.71}$ &  \cite{Aaboud:2017jvq} & $t \bar{t} h$ & $3 \ell$ & $1.8^{+0.9}_{-0.7}$ \\ \hline
\cite{Sirunyan:2018shy} & $t \bar{t} h$ & $4\ell$ & $0.9^{+2.3}_{-1.6}$ &   \cite{Aaboud:2017jvq} & $t \bar{t} h$ & $2 \ell ss$ & $1.5^{+0.7}_{-0.6}$ \\ \hline
\cite{CMS-PAS-HIG-16-042} & 0-jet DF & $WW$ & $1.30^{+0.24}_{-0.23}$ &  \cite{ATLAS-CONF-2018-004} & $gg$F & $W W$ & $1.21^{+0.22}_{-0.21}$\\ \hline
\cite{CMS-PAS-HIG-16-042} & 1-jet DF& $WW$ & $1.29^{+0.32}_{-0.27}$ & \cite{ATLAS-CONF-2018-004} & VBF & $W W$ & $0.62^{+0.37}_{-0.36}$ \\ \hline
\cite{CMS-PAS-HIG-16-042} & 2-jet DF & $WW$ & $0.82^{+0.54}_{-0.50}$ & \cite{ATLAS-CONF-2017-047} & \multicolumn{2}{c |}{B$(h \to \gamma\gamma) / $ B$(h \to 4 \ell)$} & $0.69^{+0.15}_{-0.13}$ \\ \hline
\cite{CMS-PAS-HIG-16-042} & VBF 2-jet & $WW$ & $0.72^{+0.44}_{-0.41}$ & \cite{ATLAS-CONF-2017-047} & 0-jet & $4 \ell$ & $1.07^{+0.27}_{-0.25}$ \\ \hline
\cite{CMS-PAS-HIG-16-042} & $V h$ 2-jet & $WW$ & $3.92^{+1.32}_{-1.17}$ & \cite{ATLAS-CONF-2017-047} & 1-jet, $p_T < 60$ & $4 \ell$ & $0.67^{+0.72}_{-0.68}$ \\ \hline
\cite{CMS-PAS-HIG-16-042} & $W h$ 3-lep & $WW$ & $2.23^{+1.76}_{-1.53}$ & \cite{ATLAS-CONF-2017-047} & 1-jet, $p_T \in (60,120)$ & $4 \ell$ & $1.00^{+0.63}_{-0.55}$ \\ \hline
\cite{Sirunyan:2018ouh} & $gg$F & $\gamma\gamma$ & $1.10^{+0.20}_{-0.18}$ & \cite{ATLAS-CONF-2017-047} & 1-jet, $p_T \in (120, 200)$ & $4 \ell$ & $2.1^{+1.5}_{-1.3} $\\ \hline
\cite{Sirunyan:2018ouh} & VBF & $\gamma\gamma$ & $0.8^{+0.6}_{-0.5}$ & \cite{ATLAS-CONF-2017-047} & 2-jet & $4 \ell$ & $2.2^{+1.1}_{-1.0}$ \\ \hline
\cite{Sirunyan:2018ouh} & $t \bar{t} h$ & $\gamma\gamma$ & $2.2^{+0.9}_{-0.8}$ & \cite{ATLAS-CONF-2017-047} & ``BSM-like'' & $4 \ell$ & $2.3^{+1.2}_{-1.0}$ \\ \hline
\cite{Sirunyan:2018ouh} & $V h$ & $\gamma\gamma$ & $2.4^{+1.1}_{-1.0}$ &  \cite{ATLAS-CONF-2017-047} & VBF, $p_T < 200$ & $4 \ell$ & $2.14^{+0.94}_{-0.77}$ \\ \hline
\cite{Sirunyan:2017exp} & $gg$F & $4 \ell$ & $1.20^{+0.22}_{-0.21}$ & \cite{ATLAS-CONF-2017-047} & $V h$ lep & $4 \ell$ & $0.3^{+1.3}_{-1.2}$ \\ \hline
\cite{Sirunyan:2017khh} & 0-jet & $\tau \tau$ & $0.84 \pm 0.89$ & \cite{ATLAS-CONF-2017-047} & $t \bar{t} h$ & $4 \ell$ & $0.51^{+0.86}_{-0.70}$ \\ \hline
\cite{Sirunyan:2017khh} & boosted & $\tau \tau$ & $1.17^{+0.47}_{-0.40}$ & \cite{ATLAS-CONF-2016-112} & $W h$ & $W W$ & $3.2^{+4.4}_{-4.2}$ \\ \hline
\cite{Sirunyan:2017khh} & VBF & $\tau \tau$ & $1.11^{+0.34}_{-0.35}$ & & & & \\ \hline
\cite{CMS-PAS-HIG-16-042} & $Z h$ 4-lep & $WW$ & $0.77^{+1.49}_{-1.20}$ & & & & \\ \hline
 \end{tabular}
% \end{center}
  \caption{\it Summary of LHC Run 2 Higgs results used in this work. The left side of the Table lists results from CMS,
  and the right side lists results from ATLAS.}
  \label{tab:Hrun2}
\end{table}

%%%%%%%%%%%%%%%%%%%%%%
%%%%%%%%%%%%%%%%%%%%%%
\section{Fit Methodology}
\label{sec:meth}
%%%%%%%%%%%%%%%%%%%%%%
%{\color{red} I don't know how much of this needs to go into the final version, but it is a starting point.}

We assume Gaussian errors throughout and use the method of least squares to perform our estimation of the SMEFT parameters.
The least-squares estimators for the parameters of interest, $\mathbf{\hat{c}}$, are defined by the $\chi^2$ function
\begin{equation}
\label{eq:chi2}
\chi^2\left(\mathbf{c}\right) = \left(\mathbf{y} - \boldsymbol{\mu}\left(\mathbf{c}\right)\right)^{\top} \mathbf{V}^{-1} \left(\mathbf{y} - \boldsymbol{\mu}\left(\mathbf{c}\right) \right) \, ,
\end{equation}
where the measurements tabulated in Section~\ref{sec:data} have been collected into a vector of central values, $\mathbf{y}$, 
along with a covariance matrix, $\mathbf{V}$. 
The SMEFT values of the corresponding observables have been expressed
as a vector, $\boldsymbol{\mu} = \boldsymbol{\mu}_{SM} + \mathbf{H} \cdot \mathbf{c}$, where
$\boldsymbol{\mu}_{SM}$ represents the predictions in the SM, $\mathbf{c}$ is a vector of SMEFT Wilson coefficients,
and $\mathbf{H}$ is a matrix that parameterizes in the linear approximation we use here
the SMEFT corrections to the SM predictions.

The least-squares estimators $\mathbf{\hat{c}}$ for the Wilson coefficients are found by extremizing the $\chi^2$ function, $\mathbf{w} \equiv \boldsymbol{\nabla} \chi^2 = 0$:
\begin{equation}
\label{eq:chat}
\mathbf{\hat{c}} = \left(\mathbf{H}^{\top} \mathbf{V}^{-1} \mathbf{H}\right)^{-1} \mathbf{H}^{\top} \mathbf{V}^{-1} \left(\mathbf{y} - \boldsymbol{\mu}_{SM}\right) .
\end{equation}
The covariance matrix for the least-squares estimators, $\mathbf{U}$, is given by the inverse of the Hessian of the $\chi^2$ function, $F_{ij} \equiv \tfrac{1}{2}\nabla_i \nabla_j \chi^2$:
\begin{equation}
\label{eq:U}
\mathbf{U} = \left(\mathbf{H}^{\top} \mathbf{V}^{-1} \mathbf{H}\right)^{-1} = \mathbf{F}^{-1} .
\end{equation}
The quantity in parentheses in Eq.~\eqref{eq:U} is also known as the Fisher information.
With these definitions an alternative way of writing the chi-squared function is
\begin{equation}
\label{eq:chi2alt}
\chi^2\left(\mathbf{c}\right) = \chi^2_{\text{min}} + \left(\mathbf{c} - \mathbf{\hat{c}}\right)^{\top} \cdot \mathbf{\hat{w}} + \left(\mathbf{c} - \mathbf{\hat{c}}\right)^{\top} \cdot \mathbf{F} \cdot \left(\mathbf{c} - \mathbf{\hat{c}}\right) ,
\end{equation}
where $\mathbf{\hat{w}}$ is the gradient of the chi-squared function evaluated using the least-squared estimators.

Since our analysis is to linear order in the Wilson coefficients, the likelihood associated with our $\chi^2$ 
function is a multivariate Gaussian distribution. 
As such, it is simple to compute the marginalized likelihood for a given subset of Wilson coefficients.
It is not necessary to do any integration, one simply drops the variables the one wants to marginalize 
over from $\mathbf{c},\, \hat{\mathbf{c}},\, \text{and } \mathbf{U}$.
We note also that the marginalized and profiled likelihoods for a given subset of Wilson coefficients 
are equivalent in the Gaussian approximation, which is not true in general.

%%%%%%%%%%%%%%%%%%%%%%
%%%%%%%%%%%%%%%%%%%%%%
\section{Results}
\label{sec:res}
%%%%%%%%%%%%%%%%%%%%%%
\subsection{Oblique Parameters $S$ and $T$}

As an introduction to the results from our updated global fit, we first present its implications in a simplified case where only the 
oblique parameters $\Delta S$ and $\Delta T$ introduced in~\cite{Kennedy:1988sn, Holdom:1990tc, Golden:1990ig, Altarelli:1990zd, Grinstein:1991cd, Peskin:1991sw} are non-zero.
In the Warsaw basis these parameters are given by
\begin{equation}
\frac{v^2}{\Lambda^2} C_{HWB} = \frac{g_1 g_2}{16 \pi} \Delta S, \quad \frac{v^2}{\Lambda^2} C_{HD} = - \frac{g_1 g_2}{2 \pi \left(g_1 + g_2\right)} \Delta T \, ,
\end{equation}
whereas in the SILH basis the relation (at leading order) is given by $\alpha \Delta T = \bar c_T$ and $\alpha \Delta S= 4 s_W^2 (\bar c_W + \bar c_B)$.

%%%%%%%
\begin{figure}[t!]
  \centering
\includegraphics[width=0.5\textwidth]{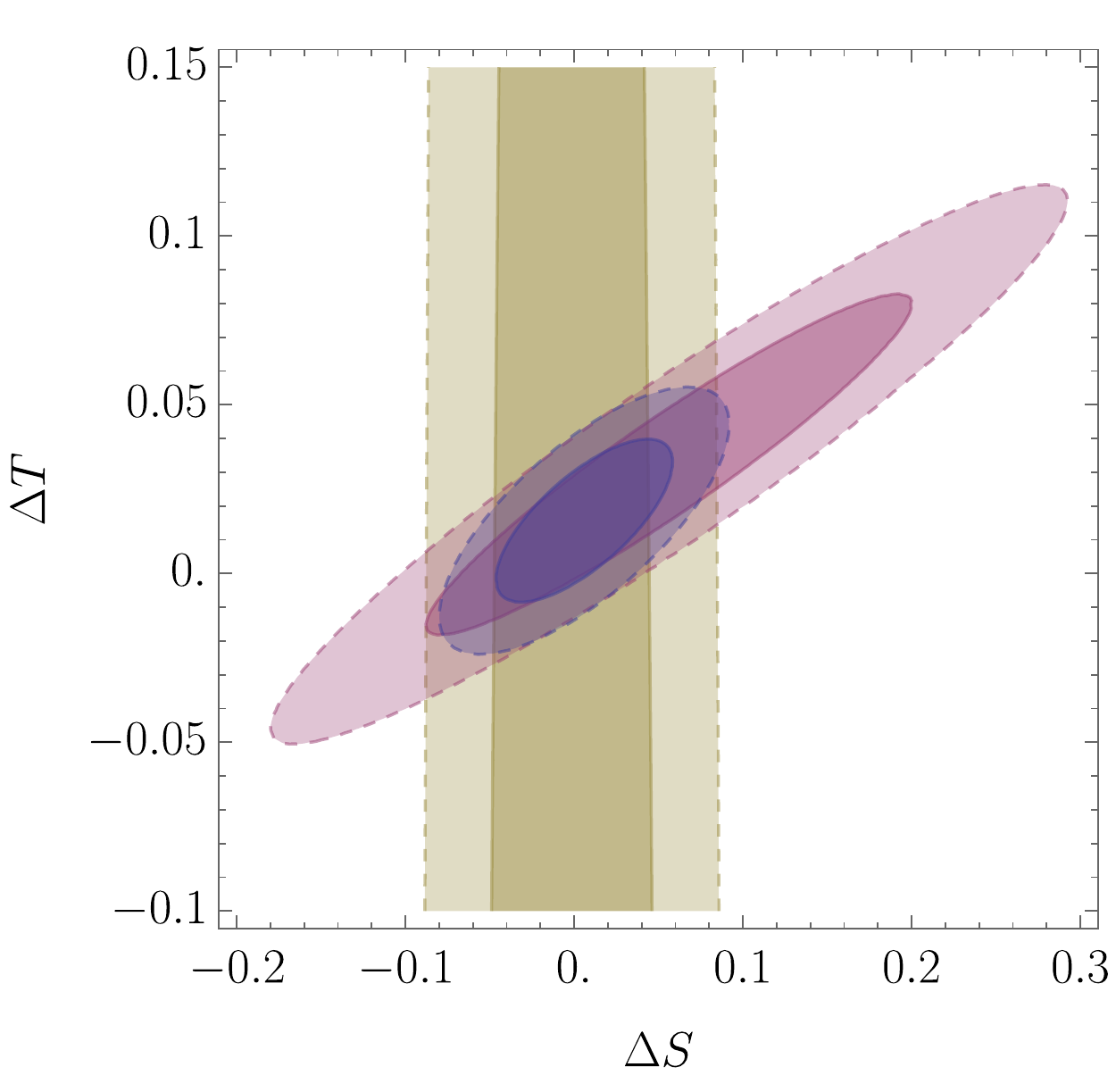}
 \caption{\it Fits to the $\Delta S$ and $\Delta T$ parameters~\protect\cite{Kennedy:1988sn, Holdom:1990tc, Golden:1990ig, Altarelli:1990zd, Grinstein:1991cd, Peskin:1991sw} using $Z$-pole, $W$ mass, and LEP 2 $WW$ scattering measurements (red), using LHC Run 1 and Run 2 Higgs results (dark yellow), and all the data (blue). The darker and lighter shaded regions are allowed at 1 and 2$\sigma$, respectively. We see that the Higgs measurements at the LHC have similar impacts to the electroweak precision measurements, 
 and are largely complementary, emphasizing the need for a combined global fit.}
   \label{fig:SandT}
\end{figure}
%%%%%%%

Figure~\ref{fig:SandT} shows the preferred parameter space for 
$\Delta S$ and $\Delta T$ for three different selections of the data sets included in the fit.
The green ellipses are obtained using just the $Z$-pole, $W$ mass, and LEP 2 $WW$ scattering measurements in the fit, 
whereas the orange ellipses use only the LHC Run 1 and Run 2 Higgs results.
Finally, the blue ellipses are obtained using all the data described in Section~\ref{sec:data}.
The regions shaded in darker and lighter  colours are allowed at 1 and 2$\sigma$, respectively. 
The 2-$\sigma$ marginalized ranges of $\Delta S$ and $\Delta T$ are 
\bea
\Delta S \in [-0.06, 0.07] , \nonumber \\
\Delta T \in [-0.02,0.05] , 
\eea
with a correlation coefficient of 0.72. 

This two-dimensional fit is restricted to the two operators in the Warsaw basis that contribute to $\Delta S$ and $\Delta T$, as defined by electroweak gauge boson propagator modifications~\footnote{These operators also induce vertex corrections that enter in the $h\gamma\gamma$ coupling.}. Nevertheless,  Figure~\ref{fig:SandT} makes it clear that the importance of the Higgs measurements at the LHC is now comparable to that of the electroweak precision measurements  for certain operators, with (basis-dependent) correlations between various measurements. Moreover, these and the Higgs constraints on $\Delta S$ and $\Delta T$ are largely complementary  in the Warsaw basis~\cite{Masso:2012eq}. This exemplifies the necessity of performing a combined global fit to precision electroweak, Higgs and diboson data, as we discuss in the rest of this Section.

%%%%%%%
\begin{figure}
  \centering
  \subfloat{\includegraphics[width=0.85\textwidth]{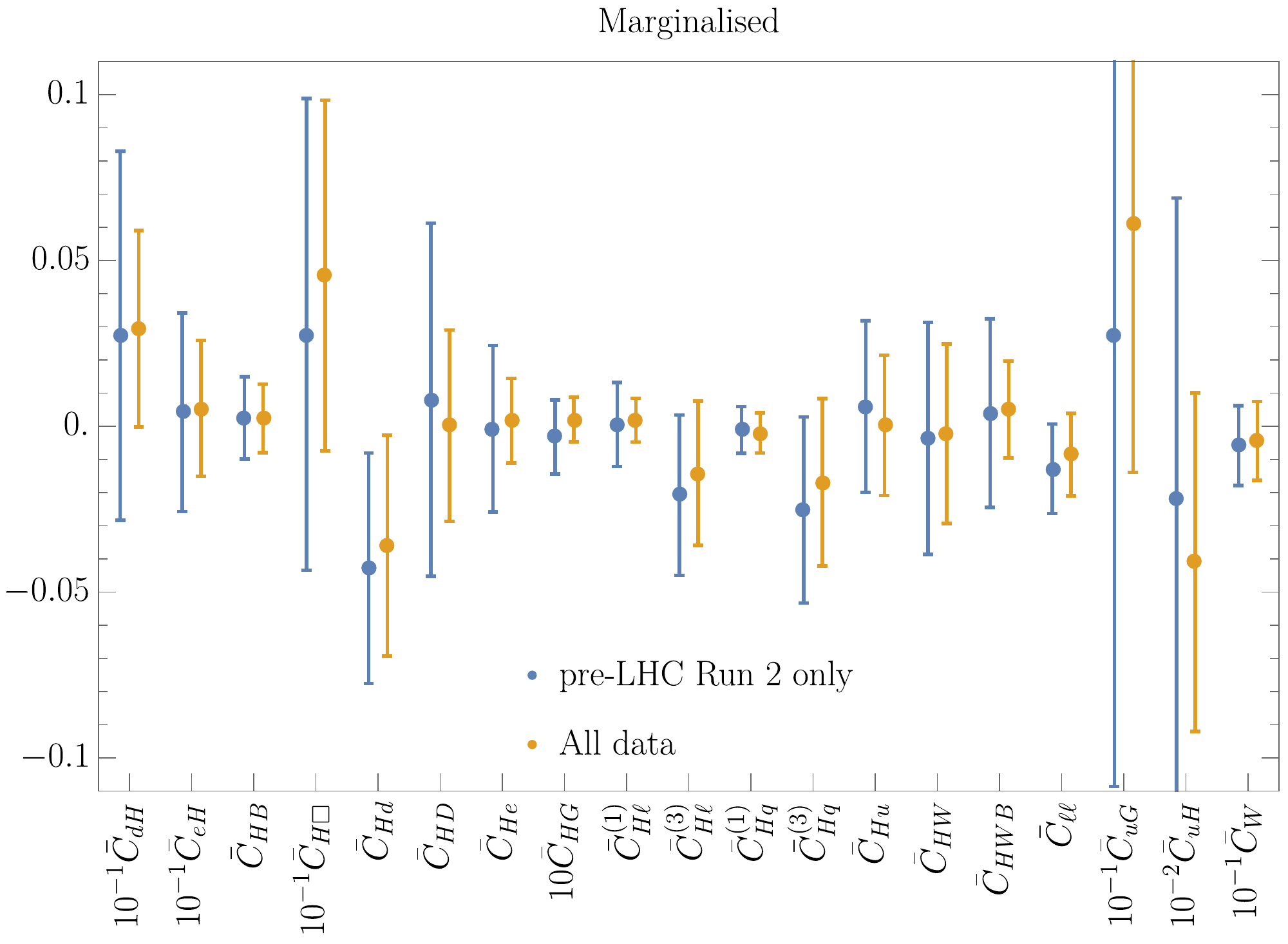}}  \\
  \subfloat{\includegraphics[width=0.85\textwidth]{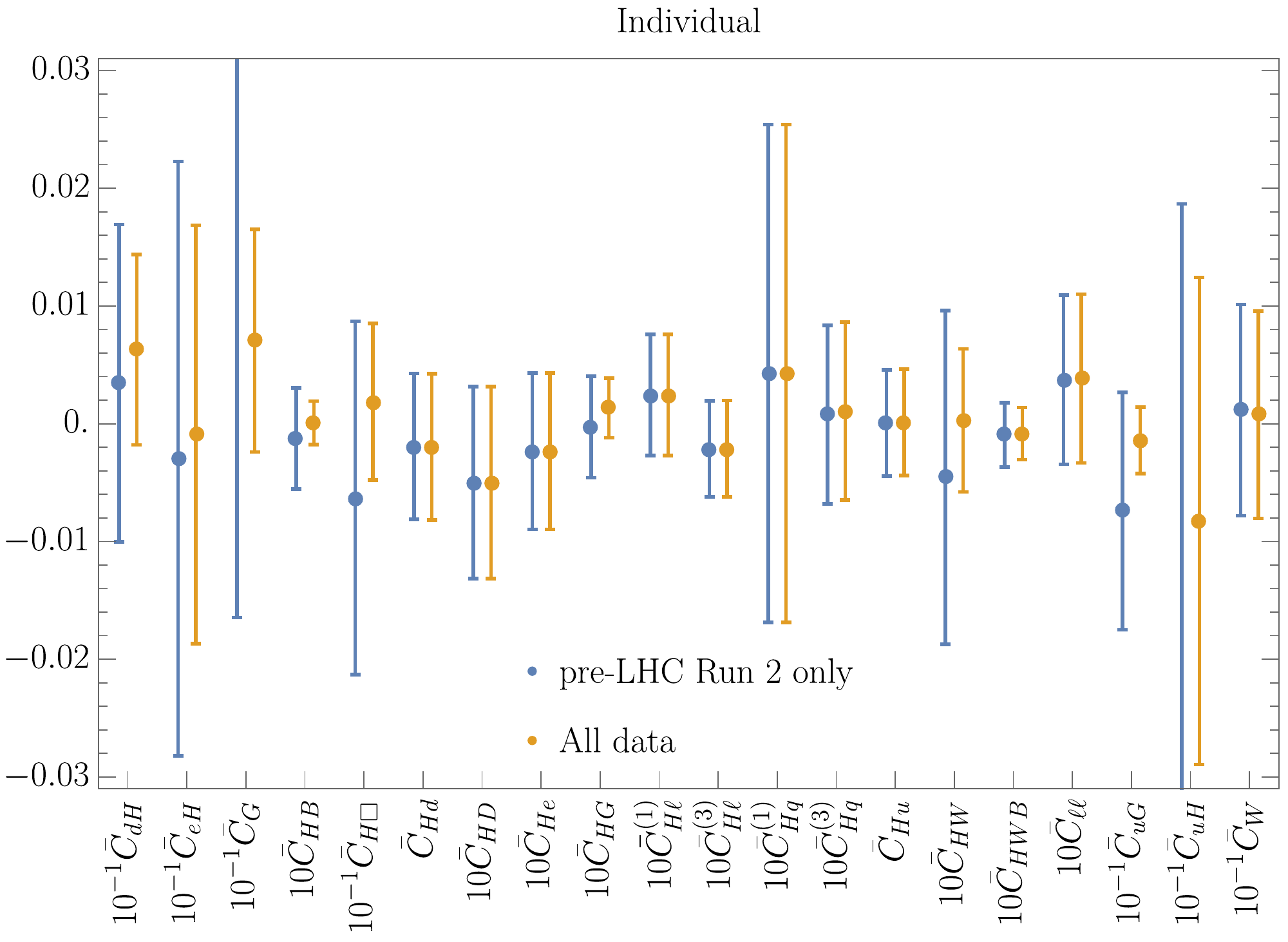}}
 \caption{\it Results from global fits in the Warsaw basis (orange) including all operators simultaneously (upper panel) and switching each operator on individually (lower panel). Also shown are fits omitting the LHC Run 2 data (blue). We display the best-fit values and 95\% CL ranges. } %JE
   \label{fig:comparison2}
\end{figure} 

%%%%%%%
\begin{figure}
\vspace{-0.5cm}
  \centering
  \subfloat{\includegraphics[width=0.83\textwidth]{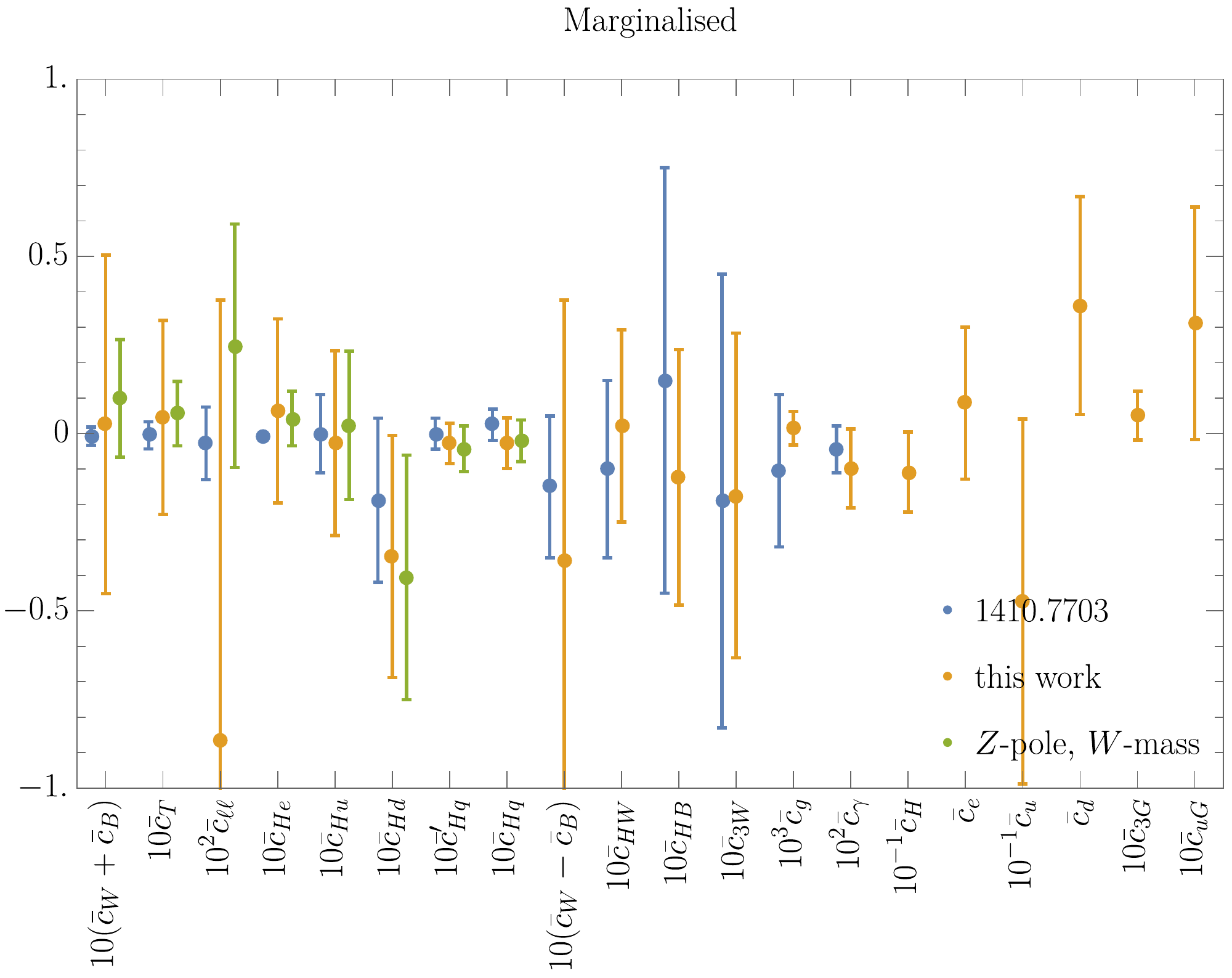}} \\
\vspace{-0.5cm}
\subfloat{\includegraphics[width=0.83\textwidth]{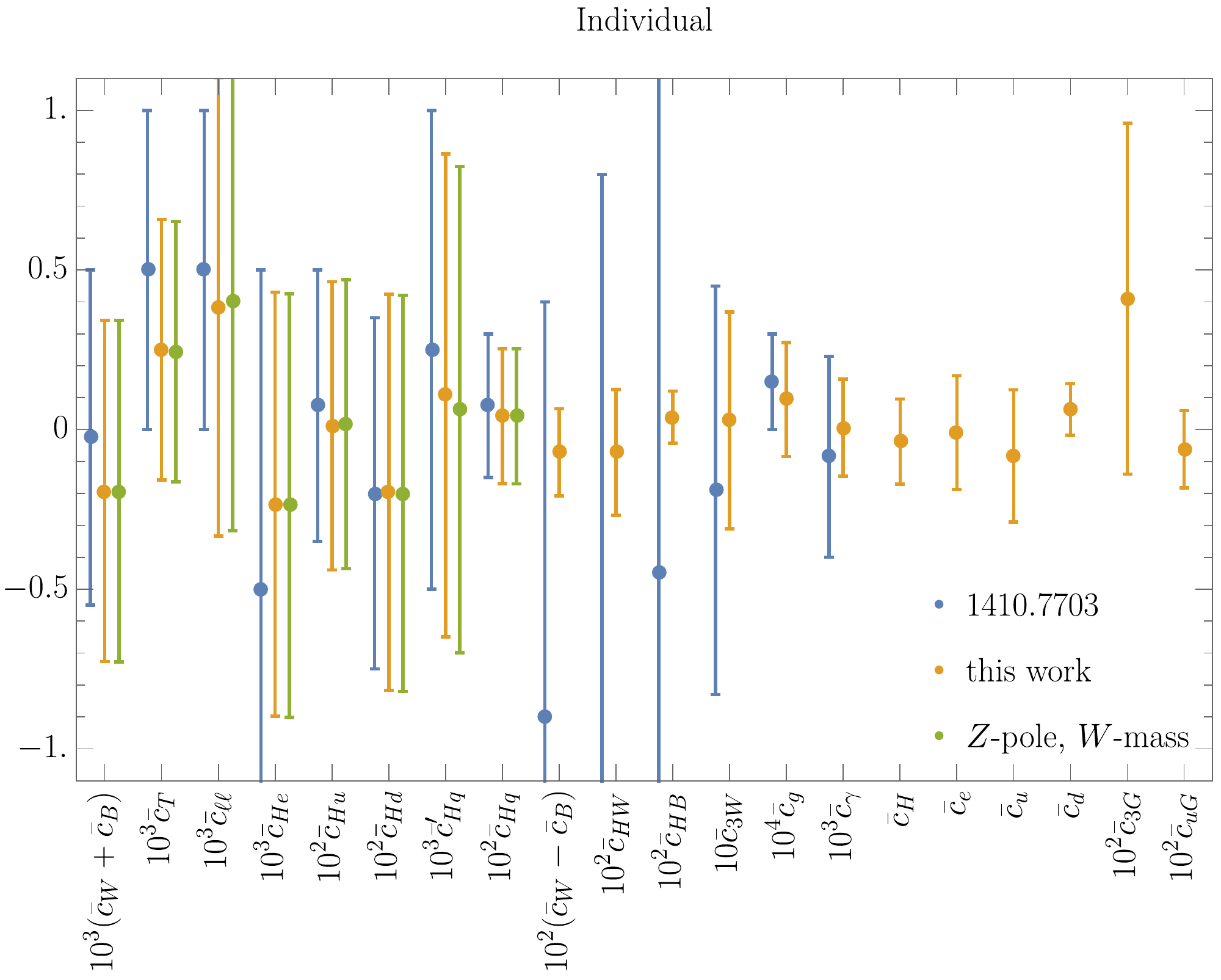}}
 \caption{\it Results from global fits in the SILH basis (orange) including all operators simultaneously (upper panel) and switching each operator on individually (lower panel). Also shown are fits to the precision electroweak $Z$-pole and $W$-mass data
 (green) and results from~\protect\cite{Ellis:2014jta} (blue). We display the best-fit values and 95\% CL ranges.} %JE
   \label{fig:comparison}
\end{figure} 

%%%%%%%%%%%%%%%%%%%%%%
\subsection{Fits to all Operator Coeficients}

With this motivation, we now turn to the results of our global fit using all the 20 dimension-6 operators discussed previously.
The upper panel of Fig.~\ref{fig:comparison2} displays our results for the best-fit values and 95\% CL ranges %JE
in the Warsaw operator basis if all these operators are included simultaneously, while the lower panel shows our results when each operator is turned on individually, with the other operator coefficients set to zero. The orange error bars are for a fit to all the measurements
described above, whereas the blue error bars are for a fit omitting the LHC Run 2 data. 
As one would expect,
the uncertainties in each operator coefficient are smaller in the fit including LHC Run 2 data, and are
generally larger in the global fit with all operators switched on than in
the fit where the operators are switched on one at a time.
The numerical results of the global fit for the 1-$\sigma$ ranges in the Warsaw basis including all sources of data %JE
are presented in the left part of Table~\ref{tab:global}.

Fig.~\ref{fig:comparison} shows the corresponding best-fit values and 95\% CL ranges in the SILH basis. %JE
The orange error bars are again for a fit to all the measurements
described above, whereas the green error bars are for a fit to the $Z$-pole and $W$ mass measurements alone. 
Again, the uncertainties in each operator coefficient are smaller when the LHC Run 2 data are included in the fit, and are
generally larger when all operators are switched on simultaneously.
The numerical results for the 1-$\sigma$ ranges in the global fit to all the available data in the SILH basis %JE
are shown in the right part of Table~\ref{tab:global}.

Fig.~\ref{fig:comparison} also compares the results of the updated global fit performed in this work
with those found in previous work in the SILH basis by three of us (JE, VS and TY) in Ref.~\cite{Ellis:2014jta}.
It should be borne in mind, when comparing the fits to see how the bounds on different coefficients have changed,
that the procedures of the two works are not identical. Nevertheless several general trends can be seen.
When considering fits to one operator at a time, the bounds on coefficients that primarily affect 
$W$- and $Z$-pole observables have not changed drastically between Ref.~\cite{Ellis:2014jta} and this work.
On the other hand, the bounds in the individual fits on the coefficients of operators that do not affect the electroweak pole observables 
have tightened, quite considerably in some cases.
When all the operators are considered simultaneously there are not such large differences
between the bounds on the operators that do not affect $W$- and $Z$- pole observables as in the one-at-a-time case.  

We show in Table~\ref{tab:impact} the relative information contents of the different sets of data for the 
different Wilson coefficients in the Warsaw basis.
A cross indicates no current sensitivity. As discussed in, e.g., Ref~\cite{Ellis:2014jta}, one can divide sets of operators in terms of their sensitivity to LEP or LHC observables.  Operators involving light fermions in the Warsaw basis had been best constrained by LEP $Z$-pole and $m_W$ constraints. The introduction of LEP $W^+ W^-$ data brings marginal gains, except for the operator $\bar c_W$ where the effect is quite dramatic. For this operator the high-energy LHC $W^+ W^-$ data do not yet improve substantially the sensitivity, although one would expect this to change as more statistics are gathered and the complete information in the full distribution is available, not just the overflow bin. 

\begin{figure}
  \centering
  \subfloat{\includegraphics[width=0.85\textwidth]{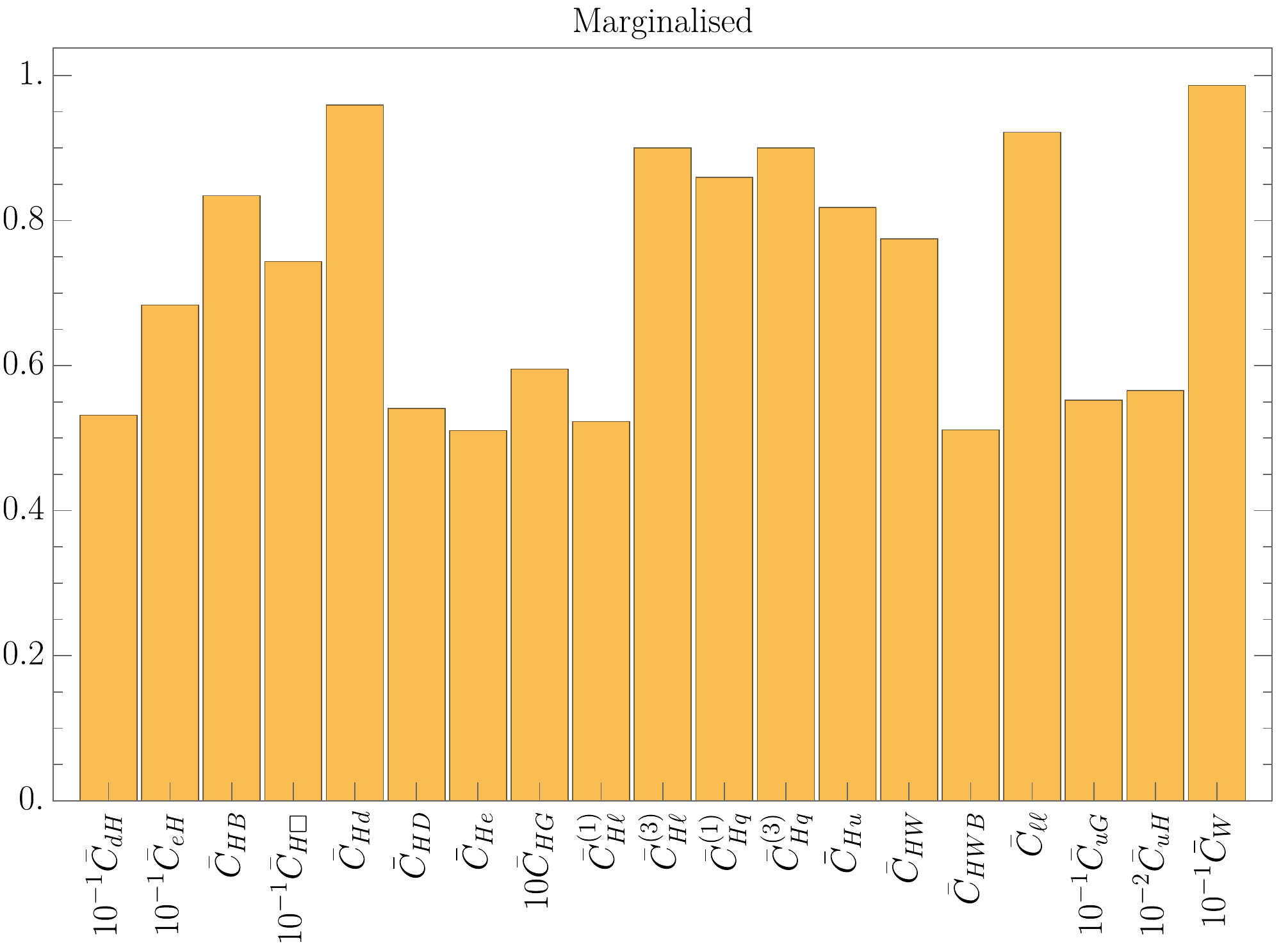}} \\
  \subfloat{\includegraphics[width=0.85\textwidth]{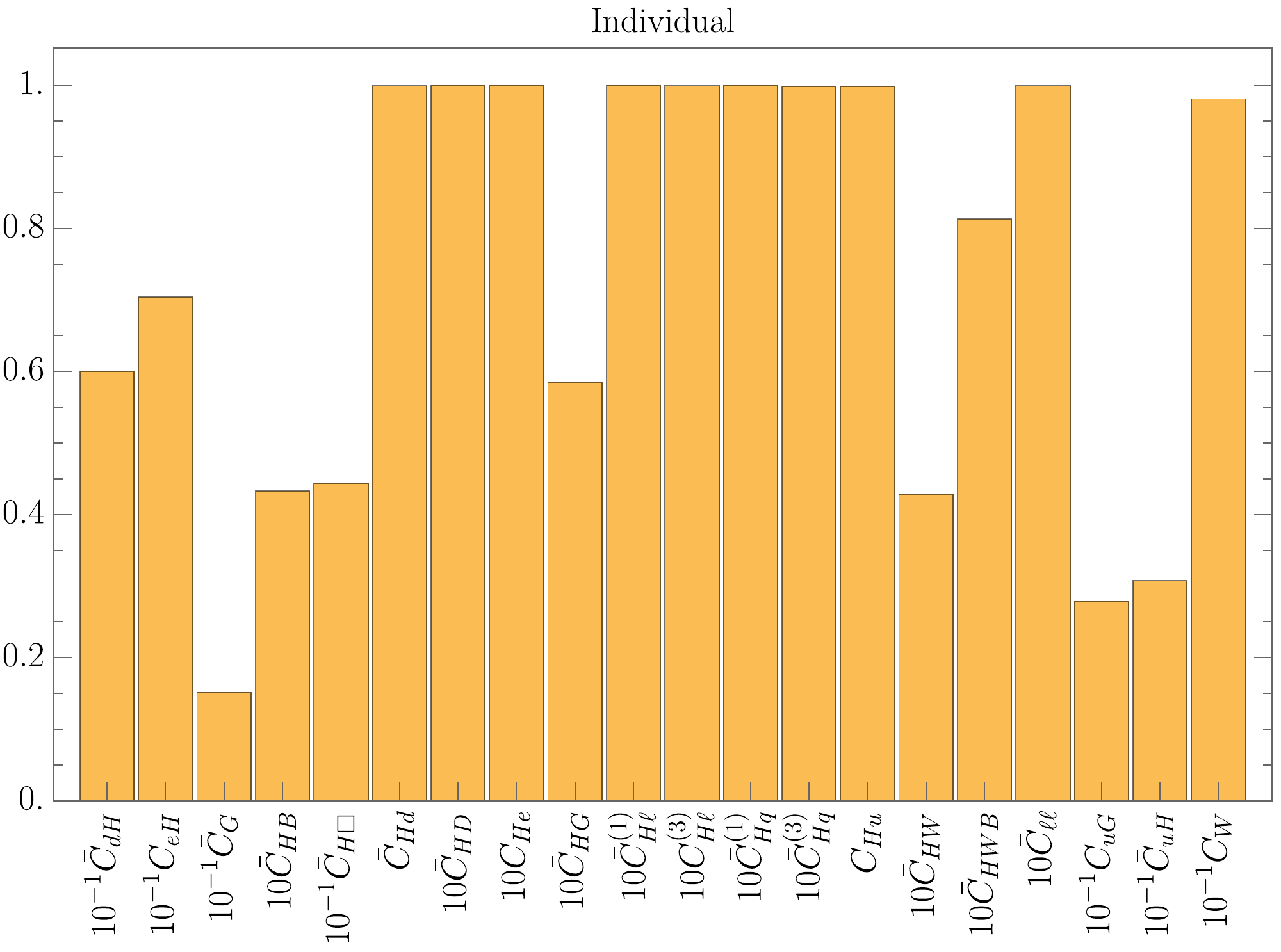}}
 \caption{\it The relative improvement in the standard deviations of the Wilson coefficients in the Warsaw basis
 when LHC Run 2 data are added to the fits (a lower number correspond to more improvement). The upper and lower panels correspond to when all operators are included simultaneously or when switching on each operator individually, respectively.}
   \label{fig:comparison3}
\end{figure}

The LHC Run~1 data opened the possibility to explore a new set of operators involving the Higgs and gauge bosons to which LEP was not sensitive. For all these operators, the Run 2 dataset is as sensitive as the Run 1 dataset, or even more sensitive. These measurements open up the sensitivity to a set of possible BSM effects that could lead to a discovery with an increased dataset in the future LHC runs.  
The relative improvements in the constraints on the Wilson coefficients in the Warsaw basis when the LHC Run 2 data
are included in the global fit are displayed graphically in Figure~\ref{fig:comparison3}. In the case where all the operators
are included (upper panel), the constraints on all the operator coefficients are improved, most significantly in the cases of $C_{HD}, C_{He}, C^{(1)}_{H \ell}$ and $C_{HWB}$, though some of the improvements are marginal, e.g., those on $C_{Hd}$ and $C_W$.
In the case where the operator coefficients are switched on individually (lower panel), the improvements in the constraints
on some coefficients are improved quite dramatically, see, e.g., $C_G$ and to a lesser extent $C_{uG}$ and $C_{uH}$,
whereas there are no improvements in the constraints on several operator coefficients, namely $C_{Hd}, C_{HD}, C_{He},
C^{(1)}_{H \ell}, C^{(3)}_{H \ell}, C^{(1)}_{Hq}, C^{(3)}_{Hq}, C_{Hu}$ and $C_{\ell \ell}$, as those are mainly constrained by electroweak precision observables. Nevertheless, we see that the improved precision of Run 2 plays an important role in improving marginalised limits. 

\begin{figure}
  \centering
  \subfloat{\includegraphics[width=0.48\textwidth]{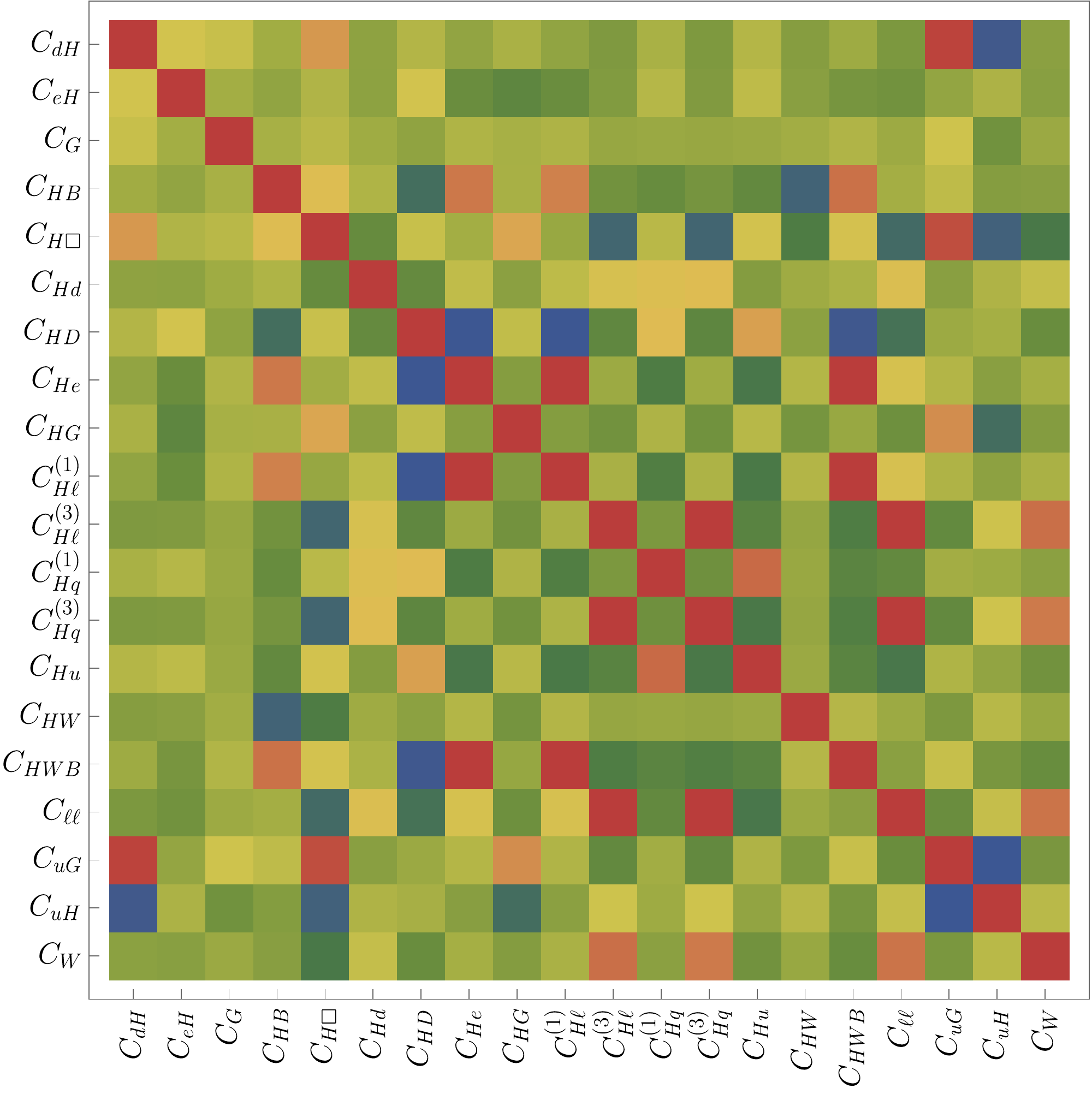}}
   \subfloat{\includegraphics[width=0.48\textwidth]{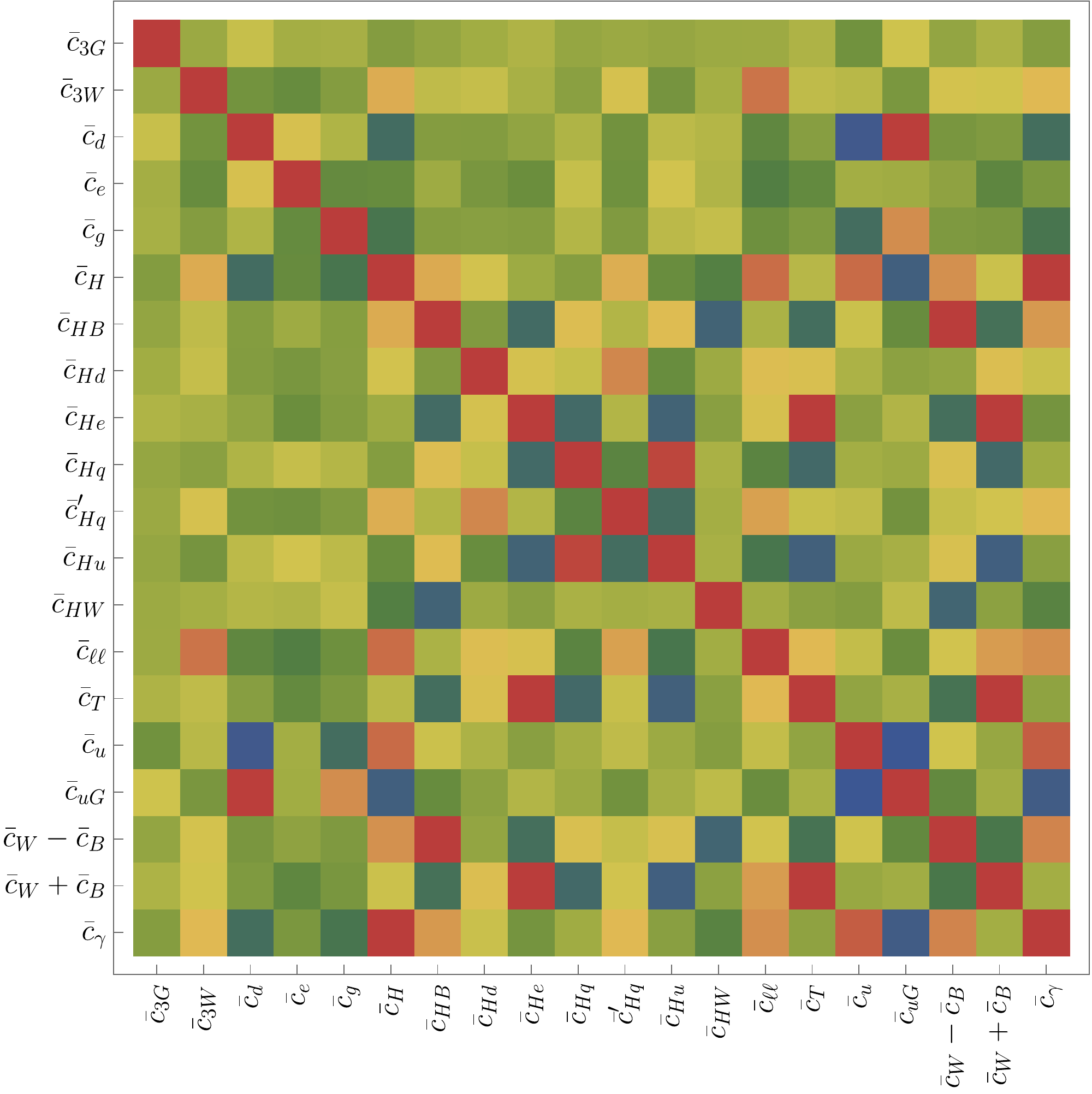}}
    \subfloat{\includegraphics[width=0.04\textwidth]{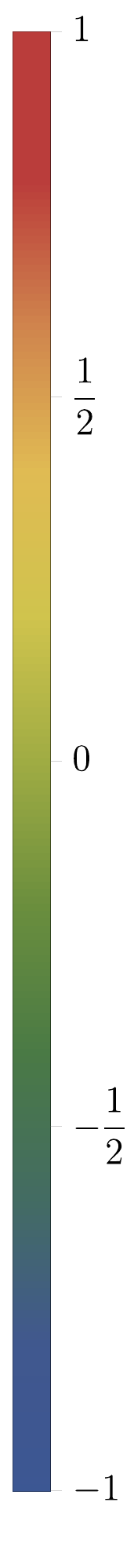}} 
 \caption{\it  Matrices of correlations among the operator coefficients in the Warsaw (left) and SILH (right) bases, as shown in Table ~\ref{tab:global}, using the colour code shown on the right. }
   \label{fig:correlations}
\end{figure}

The relative importances of these data sets are also important for the correlations between the constraints on the coefficients of the different operators. These correlations depend on the choice of basis, and we display in Figure ~\ref{fig:correlations} the correlation
matrices in the Warsaw basis (left) and the SILH basis (right), using the colour code shown in the legend on the right.
We see that both bases exhibit high degrees of correlation between some of the coefficients. In particular, in the Warsaw basis the coefficients contributing to EWPTs observables ($C_{H\ell}^{(1)}$, $C_{He}$, $C_{HD}$) as well as the pair ($C_{Hq}^{(3)}$, $C_{H\ell}^{(3)}$) are very correlated, whereas we find strong anti-correlations among operators involved in the LHC measurements ($C_G$, $C_{HW}$), ($C_{uG}$, $C_{uH}$), with operators mostly sensitive to LEP data ($C_{Hd}$, $C_{\ell\ell}$). 

On the other hand, in the SILH basis, we find strong correlations between the operators ($\bar c_{3W}$, $\bar c_{HW}$) due to the impact of diboson measurements, and correlations of the operator $\bar c_{Hq}$ with other fermionic operators $\bar c_{\ell\ell}$ and $\bar c'_{Hq}$, which are mostly constrained by LEP data, see Table~\ref{tab:impact}. As expected, the operator $\bar c_T$ is correlated with the combination of operators $\bar c_W+ \bar c_B$, as they both contribute to oblique corrections to the SM couplings~\footnote{
Numerical values of the correlation coefficients are available from 
{\tt  https://quark.phy.bnl.gov/} {\tt Digital\_Data\_Archive/SMEFT\_GlobalFit/}.}. 

\begin{center}
\begin{table}
\parbox{.1\linewidth}{
\centering
 \begin{tabular}{|c|c|c|}
 \hline 
Coefficient & Central value & 1-$\sigma$  \\  \hline  \hline
$\bar C_{dH}$ & 0.33 & 0.15 \\ \hline
 $\bar C_{eH}$ & 0.06 & 0.10 \\ \hline
 $\bar C_G$ & 0.09 & 0.06 \\ \hline
 $\bar C_{HB}$ & 0.003 & 0.005 \\ \hline
 $\bar C_{H\Box}$ & 0.50  & 0.27 \\ \hline
 $\bar C_{Hd}$ & -0.036 & 0.017 \\ \hline
 $\bar C_{HD}$ & -0.001 & 0.014 \\ \hline
 $\bar C_{He}$ & 0.002 & 0.007 \\ \hline
 $\bar C_{HG}$ & 0.0002 & 0.0003 \\ \hline
 $\bar C_{H\ell}^{(1)}$ & 0.002 & 0.003 \\ \hline
 $\bar C_{H\ell}^{(3)}$ & -0.015  & 0.011  \\ \hline
 $\bar C_{Hq}^{(1)}$ & -0.002 & 0.003 \\ \hline
 $\bar C_{Hq}^{(3)}$ & -0.017 & 0.013  \\ \hline
 $\bar C_{Hu}$ & 0.000 & 0.011 \\ \hline
 $\bar C_{HW}$ & -0.002  & 0.014 \\ \hline
 $\bar C_{HWB}$ & 0.006  & 0.007  \\ \hline
 $\bar C_{\ell\ell}$ & -0.009 & 0.006  \\ \hline
 $\bar C_{uG}$ & 0.7 & 0.4 \\ \hline
 $\bar C_{uH}$ & -4.8 & 2.6 \\ \hline
 $\bar C_{W}$ & -0.05 & 0.06 \\ \hline
 \end{tabular}
 % \caption{\it Warsaw basis.}
 }
 \hspace{6cm}
\parbox{.1\linewidth}{
\centering
  \begin{tabular}{|c|c|c|}
 \hline 
Coefficient & Central value & 1-$\sigma$ \\ \hline\hline
$\bar c_{3G}$ & 0.005 & 0.003 \\\hline
$\bar c_{3W}$ & -0.018  & 0.023 \\\hline
 $\bar c_d$ & 0.36 & 0.15 \\\hline
$\bar c_e$ & 0.09 & 0.11 \\\hline
$\bar c_g$ & 0.00002  & 0.00002 \\ \hline
$\bar c_H$ & -1.1  & 0.6 \\ \hline
$\bar c_{HB}$ & -0.013 & 0.018 \\ \hline
$\bar c_{Hd}$ & -0.035 & 0.017   \\ \hline
$\bar c_{He}$ & 0.007 & 0.013  \\ \hline
$\bar c_{Hq}$ & -0.003 & 0.004  \\ \hline
$\bar c_{Hq}^{\prime}$ & -0.003 & 0.003  \\ \hline
$\bar c_{Hu}$ & -0.03 & 0.013 \\ \hline
$\bar c_{HW}$ & 0.002 & 0.014 \\ \hline
$\bar c_{\ell\ell}$ & -0.009  & 0.006  \\ \hline
$\bar c_T$ & 0.005  & 0.013  \\ \hline
$\bar c_u$ & -4.7  & 2.6 \\ \hline
$\bar c_{uG}$ & 0.031  & 0.016 \\ \hline
$\bar c_W - \bar c_B$ & -0.04 & 0.04 \\ \hline
$\bar c_W + \bar c_B$ & 0.003  & 0.024 \\ \hline
$\bar c_\gamma$ & -0.001  & 0.0006 \\ \hline
 \end{tabular}
  }
    \caption{\it Numerical results of a global fit to all data, marginalizing over all coefficients, 
    evaluated in the Warsaw (left) and SILH (right) bases.}
   \label{tab:global}
\end{table}
\end{center}
%\end{landscape}

%\begin{landscape}
%\thispagestyle{mylandscape}
%\addtolength{\voffset}{1cm}
%\renewcommand{\arraystretch}{6}
%\footnotesize
\begin{table}
\centering
 \begin{tabular}{|c|c|c|c|c|c|} \hline
Coefficient & $Z$-pole + $m_W$ & $WW$ at LEP2 & Higgs Run1 & Higgs Run2 & LHC $WW$ high-p$_T$  \\  \hline \hline
$\bar C_{dH}$  & $\times$ & $\times$ & 36 & 64  & $\times$ \\ \hline
$\bar C_{eH}$ & $\times$ & $\times$ & 49.6      & 50.4 & $\times$ \\ \hline
 $\bar C_G$ & $\times$ & $\times$ & 2.3 & 97.7 & $\times$ \\ \hline
 $\bar C_{HB}$ & $\times$ & $\times$ & 19 & 81 & $\times$ \\ \hline
 $\bar C_{H\Box}$ & $\times$ & $\times$ & 19.7 & 80.3 & 0.01 \\ \hline
 $\bar C_{Hd}$  & 99.88 & $\times$ & 0.04 & 0.07 & $\times$ \\\hline
$\bar C_{HD}$ & 99.92 & 0.06  &  $\times$ &  $\times$ &  $\times$ \\\hline
$\bar C_{He}$ & 99.99 & 0.01  & $\times$ & $\times$ & $\times$ \\\hline
 $\bar C_{HG}$ & $\times$ & $\times$ & 34  & 66 & 0.02 \\ \hline
 $\bar C_{H\ell}^{(1)}$ & 99.97 & 0.03 &  $\times$ & $\times$ & $\times$ \\ \hline
 $\bar C_{H\ell}^{(3)}$ & 99.56 & 0.41  &  $\times$  & $\times$ & 0.01\\ \hline
 $\bar C_{Hq}^{(1)}$ & 99.98  & $\times$ & 0.01 & 0.01 & $\times$ \\ \hline
 $\bar C_{Hq}^{(3)}$ & 98.6 & 0.96 & 0.19 & 0.23  & 0.07  \\ \hline
 $\bar C_{Hu}$ & 99.5  & $\times$ & 0.2  & 0.3 & 0.04 \\ \hline
 $\bar C_{HW}$& $\times$ & $\times$ & 18 & 82  & $\times$ \\ \hline
 $\bar C_{HWB}$  & 57.9 & 0.02   & 8.2 & 33.9 & $\times$ \\ \hline
 $\bar C_{\ell\ell}$ & 99.66 & 0.32 & $\times$ & 0.01 & 0.01 \\ \hline
 $\bar C_{uG}$ & $\times$ & $\times$ & 7.8 & 92.2 & $\times$ \\  \hline
 $\bar C_{uH}$ & $\times$ & $\times$ & 9.5 & 90.5 & $\times$ \\ \hline
 $\bar C_{W}$ & $\times$ & 96.2  & $\times$ & $\times$ & 3.8  \\ \hline
 \hline
 \end{tabular}
  \caption{\it Impact of different sets of measurements on the fit to individual Wilson coefficients in the Warsaw basis
  as measured by the Fisher information contained in a given dataset for each coefficient. A cross indicates no (current) sensitivity.}
  \label{tab:impact}
\end{table}
%\end{landscape}  

%%%%%%%

 %%%%%%%

%%%%%%%%%%%%%%%%%%%%%%
\section{Implications for Extensions of the Standard Model}
\label{sec:BSM}

\subsection{Single-Parameter Models}
Ref.~\cite{deBlas:2017xtg} gave a complete dictionary in the Warsaw basis~\cite{Grzadkowski:2010es}
for new scalar bosons, vector-like fermions, and vector bosons that contribute to the dimension-six SMEFT 
operator coefficients at the tree level.
We use the notation of Ref.~\cite{deBlas:2017xtg} in what follows, unless explicitly stated otherwise.
The models that are constrained by our fit are listed in Table~\ref{tab:modeldefs}\footnote{We do not consider model $\mathcal{L}_1$, although it would be constrained by our fit, because its only interaction with the SM is through kinetic mixing with the Higgs field.}.
All of the vector-like fermion models are constrained by this dataset, whereas it constrains
only the color-singlet boson models.
It is worth noting that many of these models generate operators that are not constrained by this dataset.

%%%%%%
\begin{table}
\centering
 \begin{tabular}{| c | c | c | c | c || c | c | c | c | c |}
 \hline 
Name & Spin & $SU(3)$ & $SU(2)$ & $U(1)$ & Name & Spin & $SU(3)$ & $SU(2)$ & $U(1)$  \\ \hline \hline
$\mathcal{S}$ & 0 & 1 & 1 & 0 & $\Delta_1$ & $\tfrac{1}{2}$ & 1 & 2 & $-\tfrac{1}{2}$ \\ \hline
$\mathcal{S}_1$ & 0 & 1 & 1 & 1 & $\Delta_3$ & $\tfrac{1}{2}$ & 1 & 2 & $-\tfrac{1}{2}$ \\ \hline
$\varphi$ & 0 & 1 & 2 & $\tfrac{1}{2}$ & $\Sigma$ & $\tfrac{1}{2}$ & 1 & 3 & 0 \\ \hline
$\Xi$ & 0 & 1 & 3 & 0 & $\Sigma_1$ & $\tfrac{1}{2}$ & 1 & 3 & -1 \\ \hline
$\Xi_1$ & 0 & 1 & 3 & 1 & $U$ & $\tfrac{1}{2}$ & 3 & 1 & $\tfrac{2}{3}$ \\ \hline
$\mathcal{B}$ & 1 & 1 & 1 & 0 & $D$ & $\tfrac{1}{2}$ & 3 & 1 & $-\tfrac{1}{3}$ \\ \hline
$\mathcal{B}_1$ & 1 & 1 & 1 & 1 & $Q_1$ & $\tfrac{1}{2}$ & 3 & 2 & $\tfrac{1}{6}$ \\ \hline
$\mathcal{W}$ & 1 & 1 & 3 & 0 & $Q_5$ & $\tfrac{1}{2}$ & 3 & 2 & $-\tfrac{5}{6}$ \\ \hline
$\mathcal{W}_1$ & 1 & 1 & 3 & 1 & $Q_7$ & $\tfrac{1}{2}$ & 3 & 2 & $\tfrac{7}{6}$ \\ \hline
$N$ & $\tfrac{1}{2}$ & 1 & 1 & 0 & $T_1$ & $\tfrac{1}{2}$ & 3 & 3 & $-\tfrac{1}{3}$ \\ \hline
$E$ & $\tfrac{1}{2}$ & 1 & 1 & -1 & $T_2$ & $\tfrac{1}{2}$ & 3 & 3 & $\tfrac{2}{3}$ \\ \hline
 \end{tabular}
  \caption{\it Single-field extensions of the SM constrained by our analysis.}
  \label{tab:modeldefs}
\end{table}

We first consider renormalizable versions of the UV-complete models, with bounds on single-parameter models
being given in Table~\ref{tab:model1param2}. The total $\chi^2$ and the $\chi^2$ per degree of freedom ($\chi^2/n_{\rm d}$)
in the SM are given in the top row.
The subsequent rows show the total $\chi^2$ and the $\chi^2/n_{\rm d}$
The first set of models below the SM improve both the $\chi^2$ and the $\chi^2 / n_{\text{d}}$.
For these models we give the 1-$\sigma$ preferred range for the modulus of the
coupling squared, assuming a mass of 1~TeV, and for the mass assuming a coupling of unity.
The middle set of models improve only the $\chi^2$.
However, we note that in none of these cases is the improvement in either the $\chi^2$ or the $\chi^2 / n_{\text{d}}$ significant.
The bottom set of models improve neither the $\chi^2$ nor the $\chi^2 / n_{\text{d}}$.
For each of these models we give instead the 1-$\sigma$ upper limit on the modulus of the coupling squared, %JE
and the 1-$\sigma$ lower limit on the mass.
The bound on, or preferred range for, the mass of a particle is a better indicator  than the pull of the model
of how likely it is to be discovered at the LHC or some other future collider.  

The model named $\varphi$ in Ref.~\cite{deBlas:2017xtg} is equivalent to the Two-Higgs Doublet Model (2HDM);
see, e.g.,~\cite{Bernon:2015qea} for the corresponding 2HDM notation.
We give bounds on the Type-I 2HDM in Table~\ref{tab:model1param2}, which is 
characterized in part by having a universal modification of the SM Yukawa couplings.
Our fit is only sensitive to the product of couplings $Z_6 \cos\beta$ in the Type-I 2HDM where
\begin{equation}
\frac{v^2 Z_6}{M_{\varphi}^2} \approx \frac{1}{2} \tan\left(2\left(\beta-\alpha\right)\right) .
\end{equation}
For this reason we consider it a single-parameter model, and we do not perform a comprehensive analysis of the 2HDM. 
Furthermore, many such analyses already exist, 
both within~\cite{Ellis:2014jta, Belusca-Maito:2016dqe, Corbett:2017ieo} and 
outside~\cite{Cacchio:2016qyh, Chowdhury:2017aav,Haller:2018nnx} the EFT framework.
Lastly, note the preferred mass range for $M_{\varphi}$ in Table~\ref{tab:model1param2} assumes the product 
$Z_6 \cos\beta = -1$.

%%%%%%
\begin{table}
\centering
 \begin{tabular}{| c | c | c | c | c |}
 \hline 
Model & $\chi^2$ & $\chi^2 / n_{\text{d}}$ & Coupling & Mass / TeV   \\ \hline \hline
SM & 157 & 0.987  & - & - \\ \hline \hline
$\mathcal{S}_1$ & 156 & 0.986  & $\left|y_{\mathcal{S}_1}\right|^2 = \left(6.3 \pm 5.9\right) \cdot 10^{-3}$ & $M_{\mathcal{S}_1} = (9.0,\, 49)$ \\ \hline
$\varphi$, Type I & 156 & 0.986  & $Z_6 \cdot \cos\beta = -0.64 \pm 0.59$ & $M_{\varphi} = (0.9,\, 4.3)$ \\ \hline
$\Xi$ & 155 & 0.984  & $\left|\kappa_{\Xi}\right|^2 = \left(4.2 \pm 3.4\right) \cdot 10^{-3}$ & $M_{\Xi} = (12,\, 35)$ \\ \hline
$N$ & 155 & 0.978 & $\left|\lambda_N\right|^2 = \left(1.8 \pm 1.2\right) \cdot 10^{-2}$ & $M_N = (5.8,\, 13)$ \\ \hline
$\mathcal{W}_1$ & 155 & 0.984  & $\left|\hat{g}^{\phi}_{\mathcal{W}_1}\right|^2 = \left(3.3 \pm 2.7\right) \cdot 10^{-3}$ & $M_{\mathcal{W}_1} = (4.1,\, 13)$ \\ \hline \hline
$E$ & 157 & 0.993  & $\left|\lambda_E\right|^2 < 1.2 \cdot 10^{-2}$ & $M_E > 9.2$ \\ \hline
$\Delta_3$ & 156 & 0.990  & $\left|\lambda_{\Delta_3}\right|^2 < 1.9 \cdot 10^{-2}$ & $M_{\Delta_3} > 7.3$ \\ \hline
$\Sigma$ & 157 & 0.992  & $\left|\lambda_{\Sigma}\right|^2 < 2.9 \cdot 10^{-2}$ & $M_{\Sigma} > 5.9$ \\ \hline
$Q_5$ & 156 & 0.990  & $\left|\lambda_{Q_5}\right|^2 < 0.18$ & $M_{Q_5} > 2.4$ \\ \hline
$T_2$ & 157 & 0.992  & $\left|\lambda_{T_2}\right|^2 < 7.1 \cdot 10^{-2}$ & $M_{T_2} > 3.8$ \\ \hline \hline
$\mathcal{S}$ & 157 & 0.993 & $\left|y_{\mathcal{S}}\right|^2 < 0.32 $ & $M_{\mathcal{S}} > 1.8$ \\ \hline
$\Delta_1$ & 157 & 0.993 & $\left|\lambda_{\Delta_1}\right|^2 < 5.7 \cdot 10^{-3} $ & $M_{\Delta_1} > 13$ \\ \hline
$\Sigma_1$ & 157 & 0.993 & $\left|\lambda_{\Sigma_1}\right|^2 < 7.3 \cdot 10^{-3} $ & $M_{\Sigma_1} > 12$ \\ \hline
$U$ & 157 & 0.993 & $\left|\lambda_U\right|^2 < 2.8 \cdot 10^{-2}$ & $M_U > 6.0$ \\ \hline
$D$ & 157 & 0.993 & $\left|\lambda_D\right|^2 < 1.4 \cdot 10^{-2}$ & $M_D > 8.4$ \\ \hline
$Q_7$ & 157 & 0.993 & $\left|\lambda_{Q_7}\right|^2 < 7.7 \cdot 10^{-2}$ & $M_{Q_7} > 3.6$ \\ \hline
$T_1$ & 157 & 0.993 & $\left|\lambda_{T_1}\right|^2 < 0.13$ & $M_{T_1} > 3.0$ \\ \hline
$\mathcal{B}_1$ & 157 & 0.993 & $\left|\hat{g}^{\phi}_{\mathcal{B}_1}\right|^2 < 2.4 \cdot 10^{-3}$ & $M_{\mathcal{B}_1} > 21$ \\ \hline
 \end{tabular}
  \caption{\it 
 Single-parameter renormalizable extensions of the SM, which is included for the sake of comparison. 
  The coupling bound assumes a mass of 1~TeV, and the mass range assumes a coupling of one. 
  All bounds are at the 1$-\sigma$ level. 
  The first set of models below the SM improve both the $\chi^2$ and the $\chi^2 / n_{\text{d}}$, whereas the middle set of models only improve the $\chi^2$ (numeric values have been rounded). 
  The bottom set of models improve neither the $\chi^2$ nor the $\chi^2 / n_{\text{d}}$.
  Model $\varphi$ is the 2HDM; see the text for more discussion of this model.}
  \label{tab:model1param2}
\end{table}

\subsection{Multi-Parameter Models}

\begin{figure}
  \centering
\subfloat[$\Xi_1$]{\includegraphics[width=0.49\textwidth]{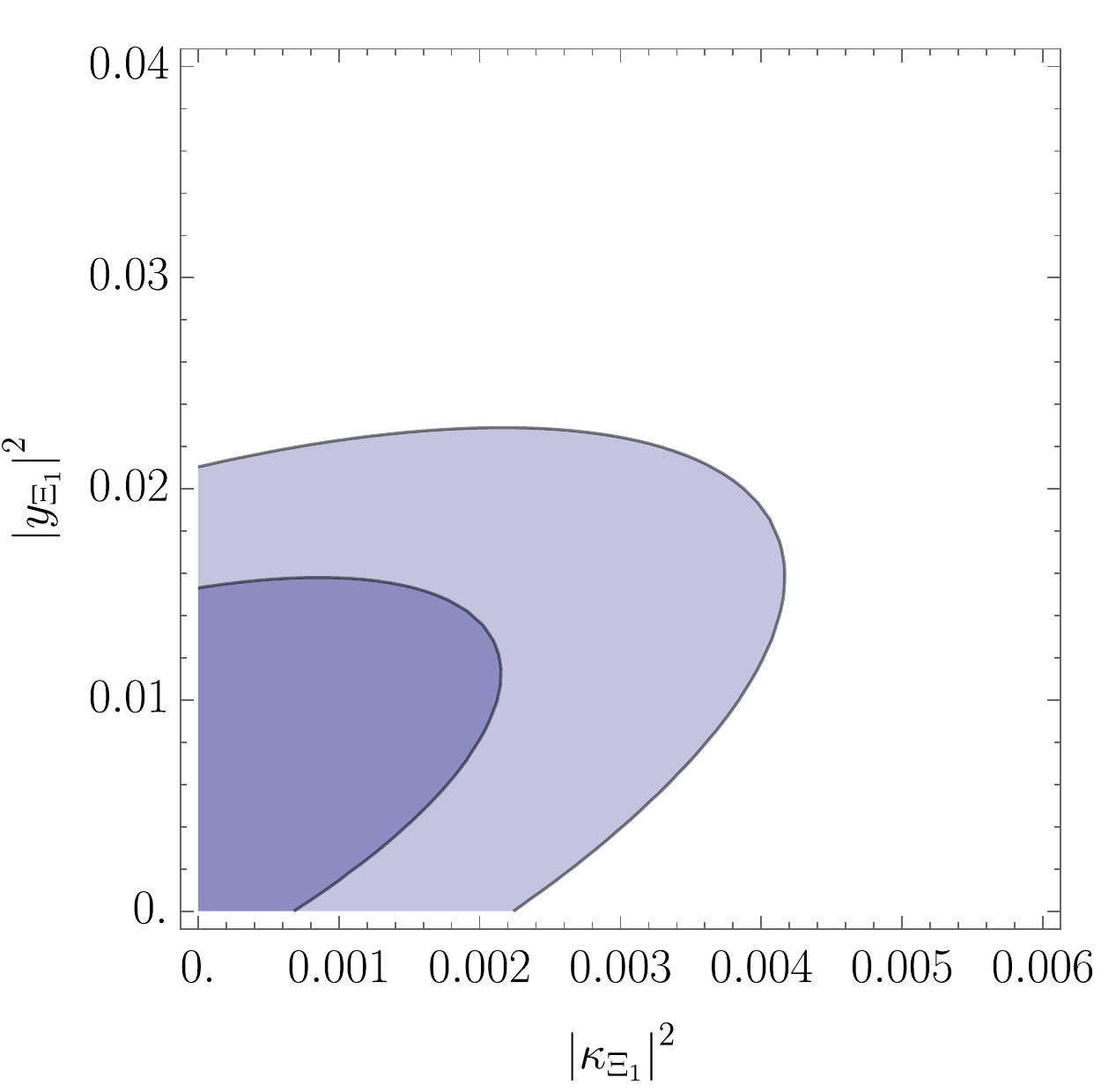}}\,
\subfloat[$Q_1$]{\includegraphics[width=0.49\textwidth]{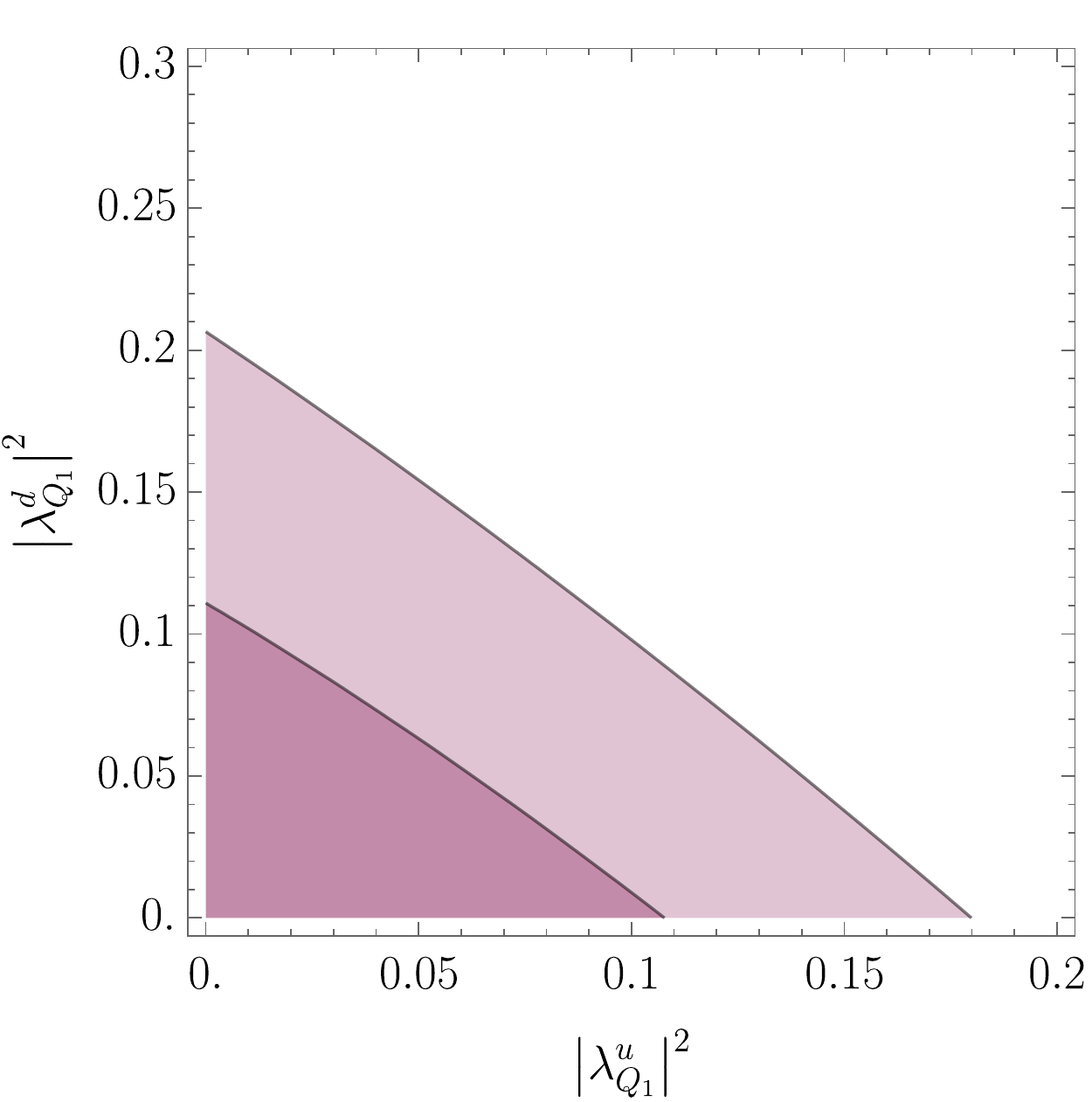}} \\
\subfloat[$\mathcal{B}$, no $\psi^4$ operators]{\includegraphics[width=0.49\textwidth]{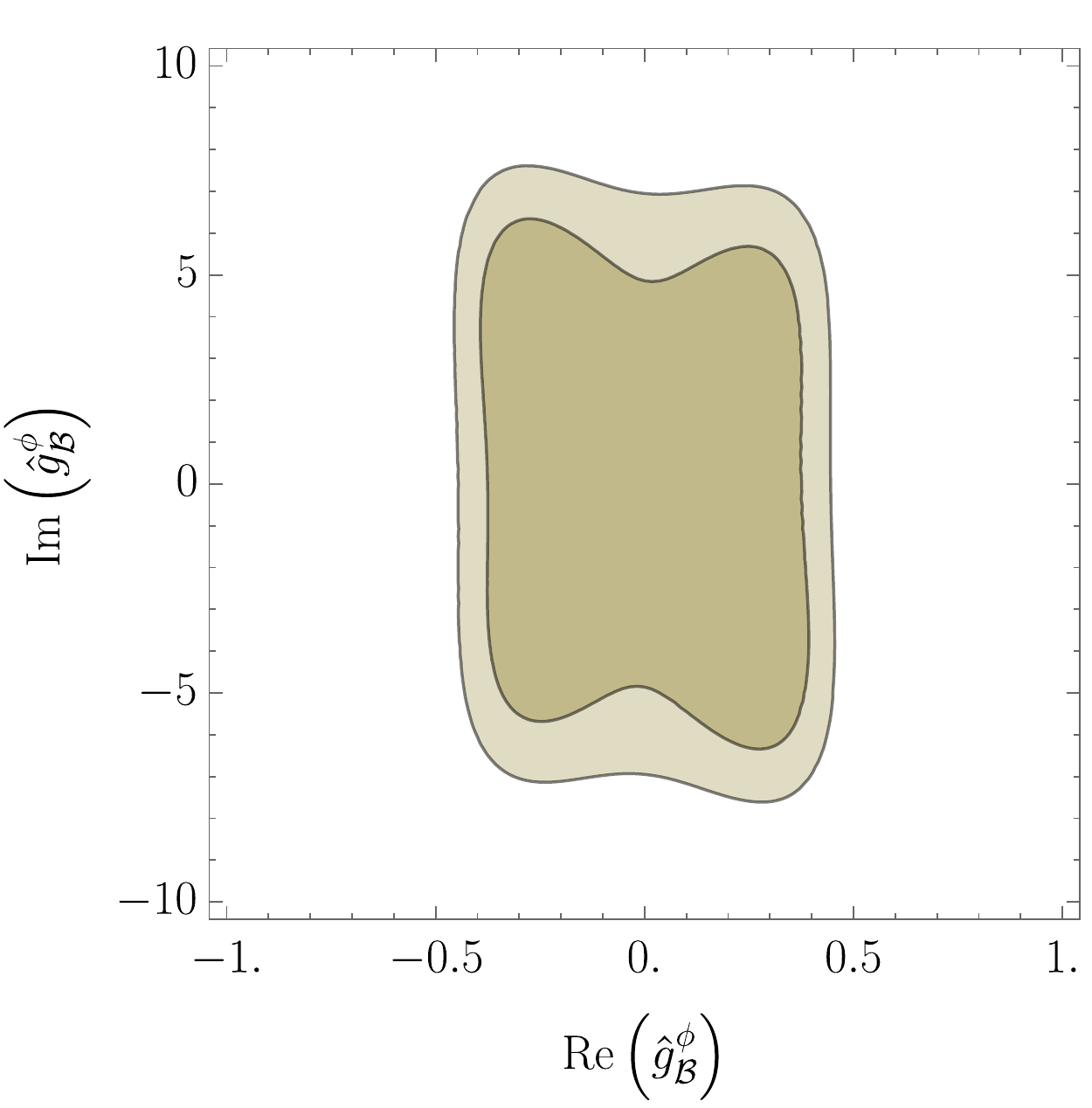}}\,
\subfloat[$\mathcal{W}$, no $\psi^4$ operators]{\includegraphics[width=0.49\textwidth]{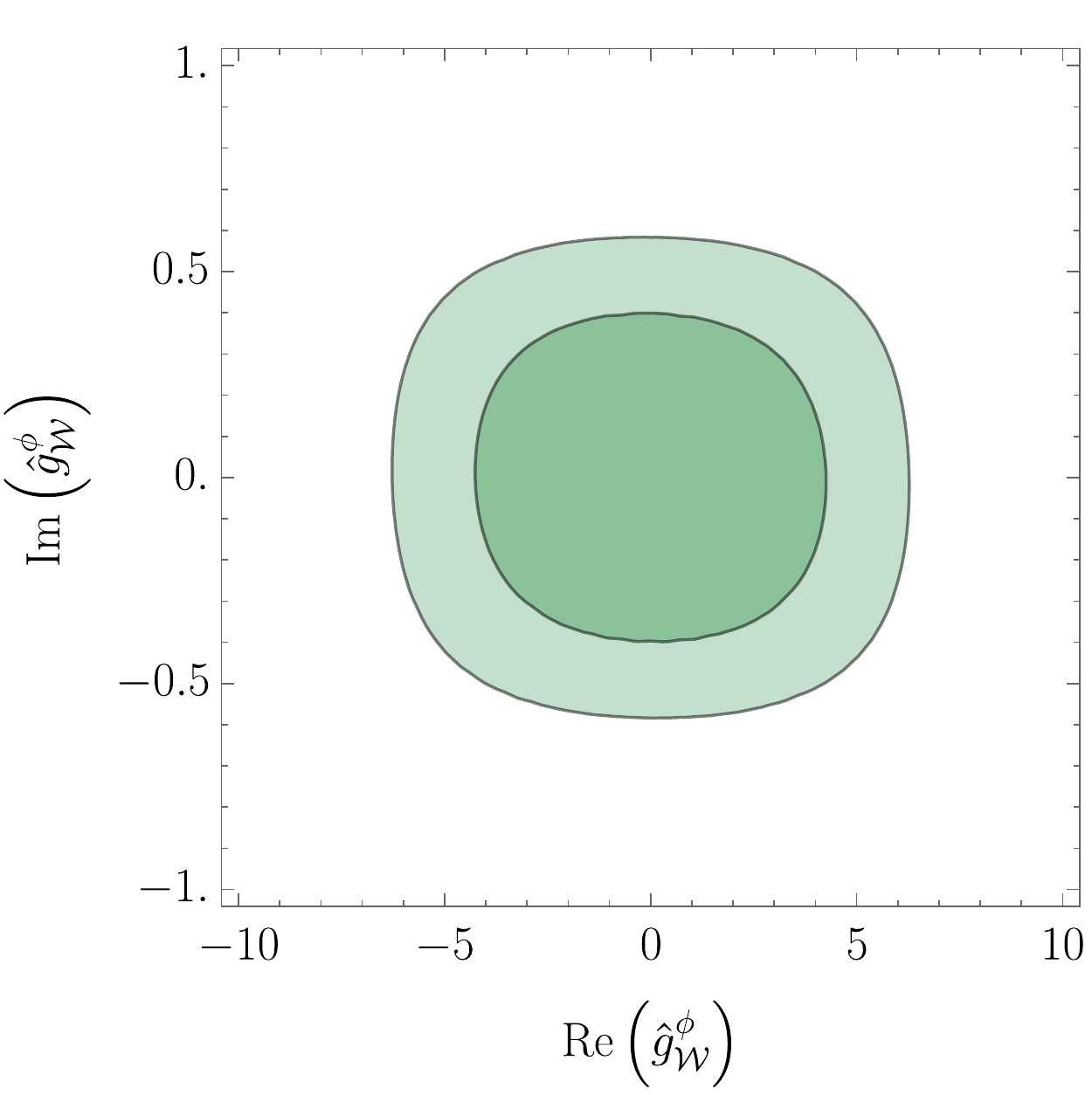}}
 \caption{\it The viable parameter spaces in the renormalizable $\Xi_1$, $Q_1$, $\mathcal{B}$, and $\mathcal{W}$ models 
 defined in Table~\protect\ref{tab:modeldefs}, assuming a mass of 1~TeV. For the two latter models we have assumed all four-fermion operators are zero. The regions shaded in darker and lighter colours are allowed at 1 and 2 $\sigma$, respectively.}
   \label{fig:model2param}
\end{figure}

We have also investigated a number of two-parameter scenarios, namely the
models $\Xi_1$, $Q_1$, $\mathcal{B}$, and $\mathcal{W}$ defined in Table~\ref{tab:modeldefs}.
For the latter two models we have assumed that all four-fermion operator coefficients are zero, %JE
both to reduce the parameter space and to avoid the bounds from dijet and dilepton searches at the LHC, 
which are not included in our fit.
The viable parameter space is each of these models assuming a mass of 1~TeV is shown in Figure~\ref{fig:model2param}.
As previously, the regions shaded in darker and lighter colours are allowed at 1 and 2 $\sigma$, respectively.

%%%%%%%

\subsection{Non-Renormalizable Models}

We now relax the assumption of renormalizability in the UV models.
In particular, dimension-5 operators are added to the UV models.
A combination of super-renormalizable and non-renormalizable operators in a UV theory can generate higher-dimensional operators with arbitrary coefficients in the corresponding low energy EFT~\cite{deBlas:2017xtg, Jenkins:2013fya}.
In a UV completion of this intermediate EFT, should it exist, these dimension-5 operators can only be generated at loop level~\cite{deBlas:2017xtg, Arzt:1994gp}.
However if this UV-completion is strongly-interacting, the coefficients generated may be sizeable, see Ref.~\cite{Manohar:2013rga} for an explicit example.

The results of fits to the non-renormalizable versions of the models in Table~\ref{tab:modeldefs} are presented in terms of the eigensystem of the covariance matrix for the least-squares estimators in Eqs.~\eqref{eq:S5}, \eqref{eq:Xi5}, \eqref{eq:Xi15}, \eqref{eq:N5}, \eqref{eq:E5}, \eqref{eq:Delta5}, \eqref{eq:Sigma5}, \eqref{eq:Sigma15}, \eqref{eq:U5}, \eqref{eq:D5}, \eqref{eq:Q15}, \eqref{eq:Q55}, \eqref{eq:Q75}, \eqref{eq:T15}, and~\eqref{eq:T25} below. 
The chi-squared and goodness-of-fit are given, as are any relations between the coefficients generated when they exist.
Note that only contributions  to the eigenvectors at the percent level or larger are presented.

\begin{itemize}
%%%
\item $\mathcal{S}^{(5)}$: $\chi^2 = 153$, $\chi^2 / n_d = 1.00$.
\begin{equation}
\label{eq:S5}
\hspace{-1.7cm}
\begin{pmatrix}
0.54 \bar{C}_{H\Box} - 0.05 \bar{C}_{HW} + 0.01 \bar{C}_{HB} + 0.08 \bar{C}_{eH} + 0.84 \bar{C}_{uH} + 0.03 \bar{C}_{dH} \\
-0.16 \bar{C}_{H\Box} + 0.75 \bar{C}_{eH} + 0.64 \bar{C}_{dH} \\
0.50 \bar{C}_{H\Box} - 0.04 \bar{C}_{HW} + 0.01 \bar{C}_{HB} + 0.57 \bar{C}_{eH} - 0.36 \bar{C}_{uH} - 0.54 \bar{C}_{dH} \\
0.65 \bar{C}_{H\Box} - 0.06 \bar{C}_{HW} + 0.02 \bar{C}_{HB} - 0.32 \bar{C}_{eH} - 0.42 \bar{C}_{uH} + 0.54 \bar{C}_{dH} \\
0.09 \bar{C}_{H\Box} + 0.95 \bar{C}_{HW} - 0.29 \bar{C}_{HB} \\
0.91 \bar{C}_{HG} + 0.12 \bar{C}_{HW} + 0.39 \bar{C}_{HB} \\
-0.39 \bar{C}_{HG} + 0.27 \bar{C}_{HW} + 0.88 \bar{C}_{HB}
\end{pmatrix}
=
\begin{pmatrix}
-0.03 \pm 0.18 \\
0.11 \pm 0.11 \\
(-4.1 \pm 7.9) \cdot 10^{-2} \\
(8.0 \pm 6.0) \cdot 10^{-2} \\
(1.8 \pm 9.6) \cdot 10^{-3} \\
(1.7 \pm 1.4) \cdot 10^{-4} \\
(2.0 \pm 8.4) \cdot 10^{-5} 
\end{pmatrix}
\end{equation}
%%%
\item $\Xi^{(5)}$: $C_{HD} = - 4 C_{H\Box}$, $\chi^2 = 152$, $\chi^2 / n_d = 0.986$. 
\begin{equation}
\label{eq:Xi5}
\begin{pmatrix}
- 0.28 \bar{C}_{eH} + 0.96 \bar{C}_{uH} - 0.04 \bar{C}_{dH} \\
0.95 \bar{C}_{eH} + 0.28 \bar{C}_{uH} + 0.14 \bar{C}_{dH} \\
-0.14 \bar{C}_{eH}  + 0.99 \bar{C}_{dH} \\
0.66 \bar{C}_{H\Box} + 0.75 \bar{C}_{HWB} \\
0.75 \bar{C}_{H\Box} - 0.66 \bar{C}_{HWB}
\end{pmatrix}
=
\begin{pmatrix}
-0.09 \pm 0.10 \\
(1.5 \pm 9.1) \cdot 10^{-2} \\
(7.3 \pm 4.5) \cdot 10^{-2} \\
(1.2 \pm 2.0) \cdot 10^{-4} \\
(8.8 \pm 8.1) \cdot 10^{-5}
\end{pmatrix}
\end{equation}
%%%
\item $\Xi_1^{(5)}$: $C_{HD} = - 4 C_{H\Box}$, $\chi^2 = 152$, $\chi^2 / n_d = 0.988$.
\begin{equation}
\label{eq:Xi15}
\begin{pmatrix}
- 0.26 \bar{C}_{eH} + 0.96 \bar{C}_{uH} - 0.03 \bar{C}_{dH} \\
0.96 \bar{C}_{eH} + 0.26 \bar{C}_{uH} + 0.08 \bar{C}_{dH} \\
-0.09 \bar{C}_{eH} + 1.0 \bar{C}_{dH} \\
- 0.19 \bar{C}_{H\Box} + 0.98 \bar{C}_{\ell\ell} \\
0.98 \bar{C}_{H\Box} - 0.19 \bar{C}_{\ell\ell}
\end{pmatrix}
=
\begin{pmatrix}
-0.09 \pm 0.10 \\
(1.9 \pm 8.9) \cdot 10^{-2} \\
(6.3 \pm 4.0) \cdot 10^{-2} \\
(1.5 \pm 4.8) \cdot 10^{-4} \\
(1.2 \pm 1.0) \cdot 10^{-4}
\end{pmatrix}
\end{equation}
%%%
\item $N^{(5)}$: $\chi^2 = 155$, $\chi^2 / n_d = 0.984$.
\begin{equation}
\label{eq:N5}
\begin{pmatrix}
0.95 \bar{C}_{H\ell}^{(1)} - 0.32 \bar{C}_{H\ell}^{(3)} \\
0.32 \bar{C}_{H\ell}^{(1)} + 0.95 \bar{C}_{H\ell}^{(3)}
\end{pmatrix}
=
\begin{pmatrix}
(3.7 \pm 2.7) \cdot 10^{-4} \\
(-1.4 \pm 2.0) \cdot 10^{-4}
\end{pmatrix}
\end{equation}
%%%
\item $E^{(5)}$: $C_{H\ell}^{(1)} = C_{H\ell}^{(3)}$, $\chi^2 = 157$, $\chi^2 / n_d = 0.999$.
\begin{equation}
\label{eq:E5} 
\begin{pmatrix}
\bar{C}_{eH} \\
\bar{C}_{H\ell}^{(3)}
\end{pmatrix}
=
\begin{pmatrix}
(-0.8 \pm 8.9) \cdot 10^{-2} \\
(-0.3 \pm 1.5) \cdot 10^{-4}
\end{pmatrix}
\end{equation}
%%%
\item $\Delta_{1,3}^{(5)}$: $\chi^2 = 156$, $\chi^2 / n_d = 0.996$.
\begin{equation}
\label{eq:Delta5}
\begin{pmatrix}
\bar{C}_{eH} \\
\bar{C}_{He}
\end{pmatrix}
=
\begin{pmatrix}
(-0.8 \pm 8.9) \cdot 10^{-2} \\
(-2.3 \pm 3.3) \cdot 10^{-4}
\end{pmatrix}
\end{equation}
%%%
\item $\Sigma^{(5)}$: $C_{H\ell}^{(1)} = 3 C_{H\ell}^{(3)}$, $\chi^2 = 157$, $\chi^2 / n_d = 0.998$.
\begin{equation}
\label{eq:Sigma5}
\begin{pmatrix}
\bar{C}_{eH} \\
\bar{C}_{H\ell}^{(3)}
\end{pmatrix}
=
\begin{pmatrix}
(-0.8 \pm 8.9) \cdot 10^{-2} \\
(3.3 \pm 7.4) \cdot 10^{-5}
\end{pmatrix}
\end{equation}
%%%
\item $\Sigma_1^{(5)}$: $C_{H\ell}^{(1)} = - 3 C_{H\ell}^{(3)}$, $\chi^2 = 155$, $\chi^2 / n_d = 0.988$.
\begin{equation}
\label{eq:Sigma15}
\begin{pmatrix}
\bar{C}_{eH} \\
\bar{C}_{H\ell}^{(3)}
\end{pmatrix}
=
\begin{pmatrix}
(-0.8 \pm 8.9) \cdot 10^{-2} \\
(-1.2 \pm 0.9) \cdot 10^{-4}
\end{pmatrix}
\end{equation}
%%%
\item $U^{(5)}$: $C_{Hq}^{(1)} = - C_{Hq}^{(3)}$, $\chi^2 = 155$, $\chi^2 / n_d = 0.993$.
\begin{equation}
\label{eq:U5}
\begin{pmatrix}
0.99 \bar{C}_{uH} - 0.13 \bar{C}_{uG} \\
0.13 \bar{C}_{uH} + 0.99 \bar{C}_{uG} \\
\bar{C}_{Hq}^{(3)}
\end{pmatrix}
=
\begin{pmatrix}
0.51 \pm 0.52 \\
(-1.4 \pm 1.4) \cdot 10^{-2} \\
(1.0 \pm 5.1) \cdot 10^{-4}
\end{pmatrix}
\end{equation}
%%%
\item $D^{(5)}$: $C_{Hq}^{(1)} = C_{Hq}^{(3)}$, $\chi^2 = 154$, $\chi^2 / n_d = 0.983$.
\begin{equation}
\label{eq:D5}
\begin{pmatrix}
\bar{C}_{dH} \\
\bar{C}_{Hq}^{(3)}
\end{pmatrix}
=
\begin{pmatrix}
(6.4 \pm 4.0) \cdot 10^{-2} \\
(1.0 \pm 2.9) \cdot 10^{-4}
\end{pmatrix}
\end{equation}
%%%
\item $Q_1^{(5)}$: $\chi^2 = 152$, $\chi^2 / n_d = 0.987$.
\begin{equation}
\label{eq:Q15}
\begin{pmatrix}
0.99 \bar{C}_{uH} - 0.07 \bar{C}_{dH} - 0.14 \bar{C}_{uG} \\
0.08 \bar{C}_{uH} + 1.0 \bar{C}_{dH} + 0.05 \bar{C}_{uG} \\
0.13 \bar{C}_{uH} - 0.06 \bar{C}_{dH} + 0.99 \bar{C}_{uG} \\
0.57 \bar{C}_{Hu} + 0.82 \bar{C}_{Hd} \\
0.82 \bar{C}_{Hu} - 0.57 \bar{C}_{Hd}
\end{pmatrix}
=
\begin{pmatrix}
-0.8 \pm 1.2 \\
(5.8 \pm 4.1) \cdot 10^{-2} \\
(-1.5 \pm 1.4) \cdot 10^{-2} \\
(-1.0 \pm 0.8) \cdot 10^{-2} \\
(0.5 \pm 1.9) \cdot 10^{-3}
\end{pmatrix}
\end{equation}
%%%
\item $Q_5^{(5)}$: $\chi^2 = 154$, $\chi^2 / n_d = 0.982$.
\begin{equation}
\label{eq:Q55}
\begin{pmatrix}
\bar{C}_{dH} \\
\bar{C}_{Hd}
\end{pmatrix}
=
\begin{pmatrix}
(6.4 \pm 4.0) \cdot 10^{-2} \\
(-2.0 \pm 3.1) \cdot 10^{-3}
\end{pmatrix}
\end{equation}
%%%
\item $Q_7^{(5)}$: $\chi^2 = 156$, $\chi^2 / n_d = 0.995$,
\begin{equation}
\label{eq:Q75}
\begin{pmatrix}
\bar{C}_{uH} \\
\bar{C}_{Hu}
\end{pmatrix}
=
\begin{pmatrix}
- 0.08 \pm 0.10 \\
(0.01 \pm 2.3) \cdot 10^{-3}
\end{pmatrix}
\end{equation}
%%%
\item $T_1^{(5)}$: $C_{Hq}^{(1)} = - 3 C_{Hq}^{(3)}$, $\chi^2 = 154$, $\chi^2 / n_d = 0.986$.
\begin{equation}
\label{eq:T15}
\begin{pmatrix}
\bar{C}_{uH} - 0.01 \bar{C}_{dH} \\
0.01 \bar{C}_{uH} + \bar{C}_{dH} \\
\bar{C}_{Hq}^{(3)}
\end{pmatrix}
=
\begin{pmatrix}
- 0.09 \pm 0.10 \\
(6.4 \pm 4.0) \cdot 10^{-2} \\
(-1.4 \pm 5.9) \cdot 10^{-4}
\end{pmatrix}
\end{equation}
%%%
\item $T_2^{(5)}$: $C_{Hq}^{(1)} = 3 C_{Hq}^{(3)}$, $\chi^2 = 154$, $\chi^2 / n_d = 0.985$.
\begin{equation}
\label{eq:T25}
\begin{pmatrix}
\bar{C}_{uH} - 0.01 \bar{C}_{dH} \\
0.01 \bar{C}_{uH} + \bar{C}_{dH} \\
\bar{C}_{Hq}^{(3)}
\end{pmatrix}
=
\begin{pmatrix}
- 0.09 \pm 0.10 \\
(6.4 \pm 4.0) \cdot 10^{-2} \\
(0.7 \pm 1.9) \cdot 10^{-4}
\end{pmatrix}
\end{equation}
\end{itemize}

\subsection{Stop Squarks in the MSSM}

Finally, as an example how the constraints on the SMEFT coefficients can be used to constrain possible BSM physics
at the loop level, we consider the minimal supersymmetric extension of the SM (the MSSM). 
Among the sparticles for which the data may be
most constraining are the stops, by virtue of their large couplings to the Higgs field. Moreover, SMEFT constraints are of particular
interest for stops also because the constraints from direct searches are model-dependent and often require
the understanding of complicated final states to which the LHC has reduced sensitivity, whereas the SMEFT constraints
are relatively  model-independent. Run 1 LHC data were used to constrain degenerate stops in~\cite{Henning:2014wua,Henning:2014gca}, and non-degenerate stops in~\cite{Drozd:2015kva}, where comparisons were made between
the constraints obtained using the SMEFT and  an exact one-loop calculation. 
It was found there that the SMEFT and exact one-loop results were quite similar, except in regions of
parameter space where the data were insensitive even to very light stops.%{\bf Accuracy of SMEFT?}
%%%%%%%
\begin{figure}[t!]
  \centering
\includegraphics[width=0.48\textwidth]{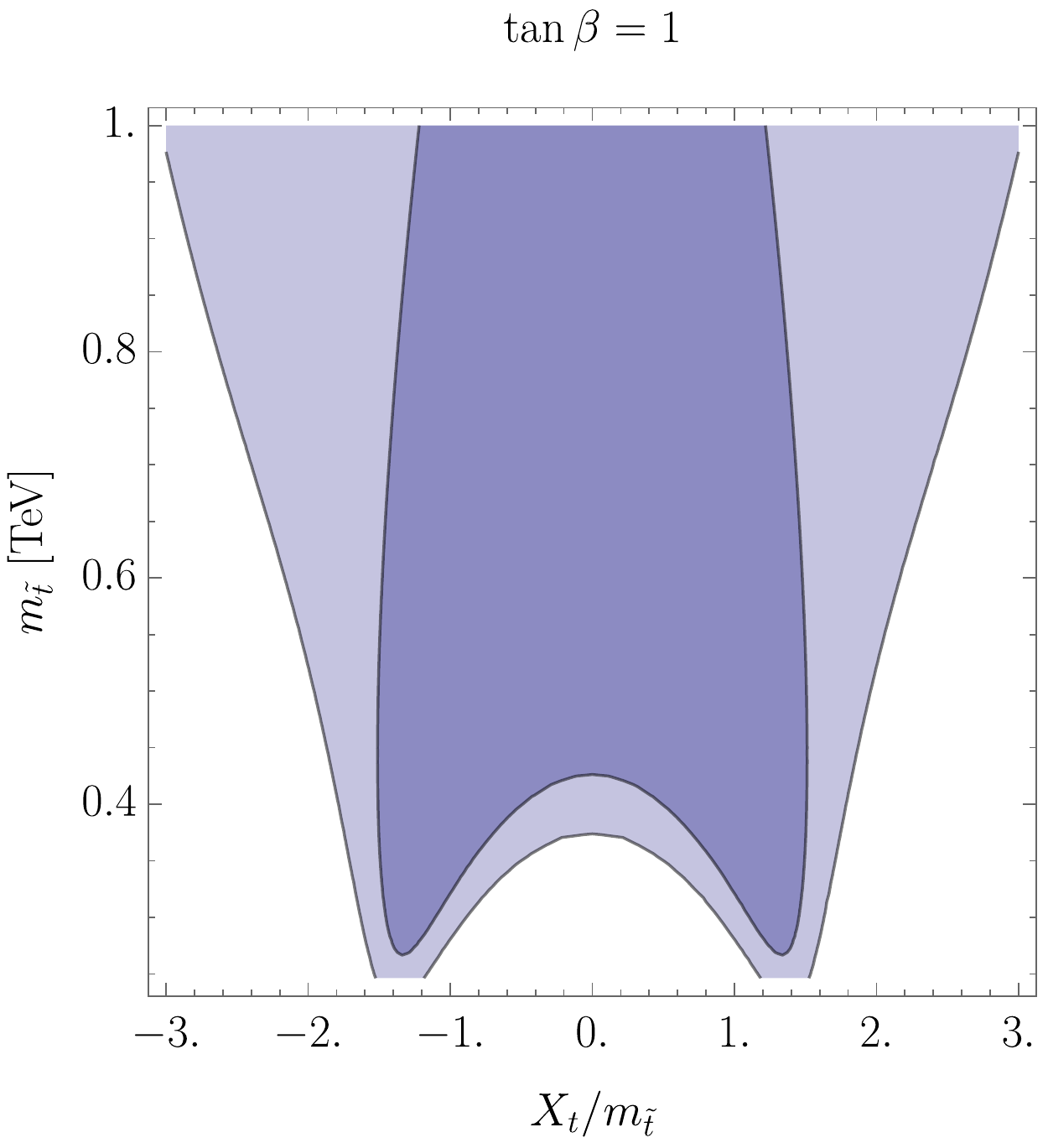}
\includegraphics[width=0.48\textwidth]{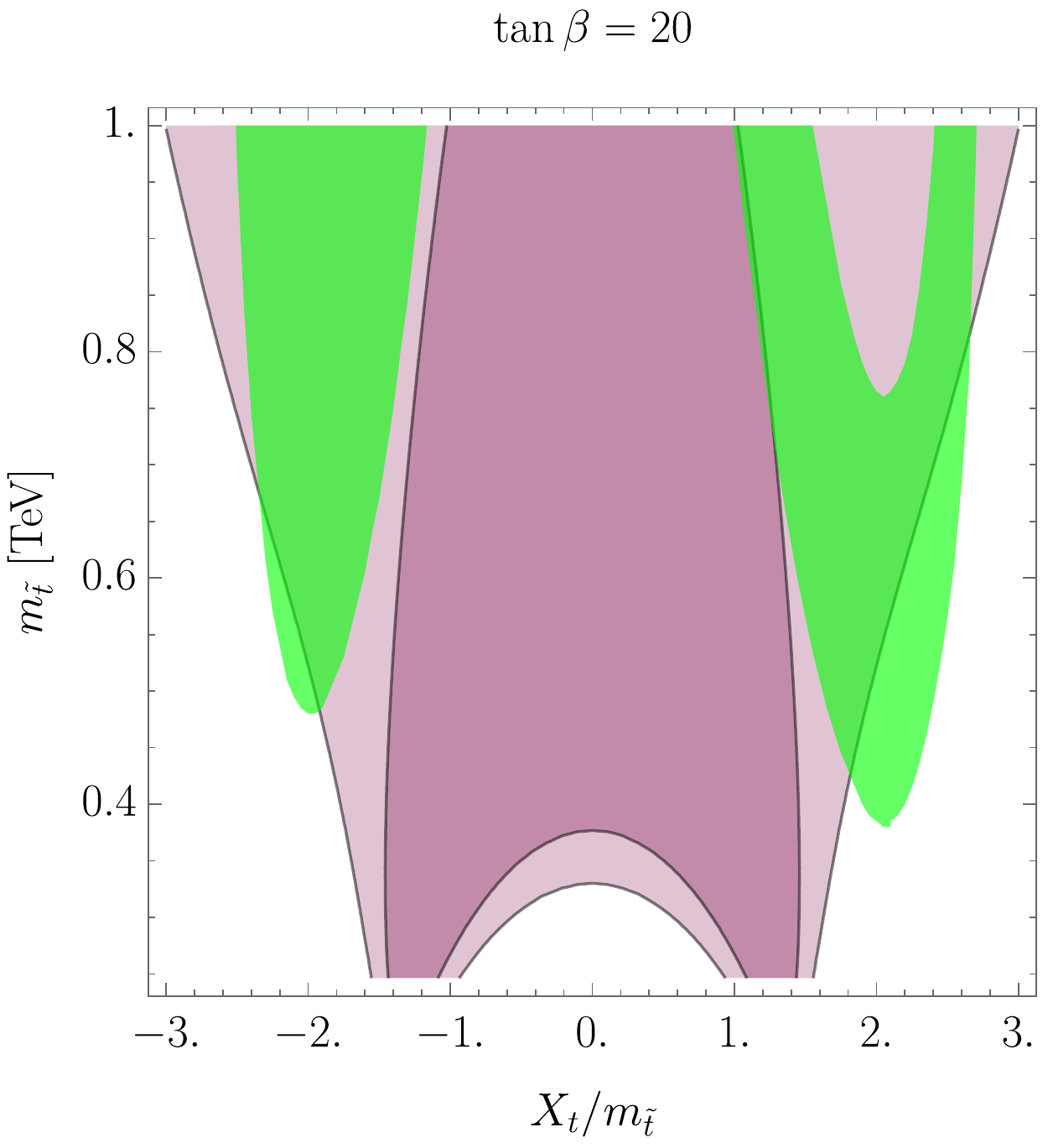}
 \caption{\it The allowed degenerate stop parameter space with $\tan\beta = 1$ (left) and $\tan\beta = 20$ (right), where the darker and lighter
 blue regions are within 1- and 2-$\sigma$ of the minimum of the $\chi^2$ function, respectively.
 In addition, the green shading indicates the region where $M_h \in (122,\,128)~\textnormal{GeV}$~\cite{Drozd:2015kva}.}
   \label{fig:stops}
\end{figure} 

Here we make a new comparison of the degenerate stop case using Run 2 LHC data.
Figure~\ref{fig:stops} shows the SMEFT constraints in the plane of the degenerate stop mass $m_{\tilde t}$ 
and the stop mixing parameter $X_t$ for the two choices $\tan\beta = 1$ and 20 of the ratio of Higgs vev's, 
where the darker and lighter blue regions correspond to 1- and 2-$\sigma$  ranges, respectively.
We note that the kinematic ranges of the LHC Run 2 Higgs data used in our analysis extend typically
to $p_T \lesssim 200$~GeV (see Table~\ref{tab:Hrun2}). The LHC $W^+ W^-$ data that we use include a tail that may extend to
higher $p_T$, but this has less weight in the global fit, see the last column in Table~\ref{tab:impact}. We therefore expect
the SMEFT analysis to be reasonably reliable for $m_{\tilde t} \gtrsim 300$~GeV.

By way of comparison, although the LHC limits on the stop mass may extend as far as $m_{\tilde t} \simeq 1$~TeV
under certain assumptions on the sparticle spectrum~\cite{wiki:ATLAS,wiki:CMS}, they are sensitive to the value assumed for the lightest supersymmetric
particle ${\tilde \chi}_0^1$, disappearing entirely for $m_{{\tilde \chi}_0^1} \lesssim 400$~GeV and having holes for some values
of $m_{\tilde t} \gtrsim 300$~GeV when $m_{{\tilde \chi}_0^1} \gtrsim 250$~GeV. We conclude that the indirect SMEFT constraint is
highly competitive, despite the fact that the Wilson coefficients are generated only at the loop level.

%%%%%%%%%%%%%%%%%%%%%%
%%%%%%%%%%%%%%%%%%%%%%
\section{Conclusions}
\label{sec:con}
%%%%%%%%%%%%%%%%%%%%%%

We have presented in this paper a first combined global analysis within the SMEFT of the available precision
electroweak data, diboson production data from LEP and the LHC, and the data on Higgs production 
from Runs 1 and 2 of the LHC. Our analysis takes into account all the 20 dimension-6 operators
that are relevant to these processes. We emphasize that these data should be analyzed jointly,
as the constraints from different data categories are synergistic, complementary and of comparable importance. This point is
exemplified in Fig.~\ref{fig:SandT}, where we see explicitly the complementarity of the constraints 
from $Z$-pole, $W$ mass, and LEP 2 $W^+ W^-$ production measurements (orange) and LHC
Higgs production measurements (green) on the oblique
parameters $S$ and $T$, which are proportional to the dimension-6 operator coefficients
$C_{HWB}$ and $C_{HD}$, respectively.

\begin{figure}[t!]
  \centering
\includegraphics[width=0.8\textwidth]{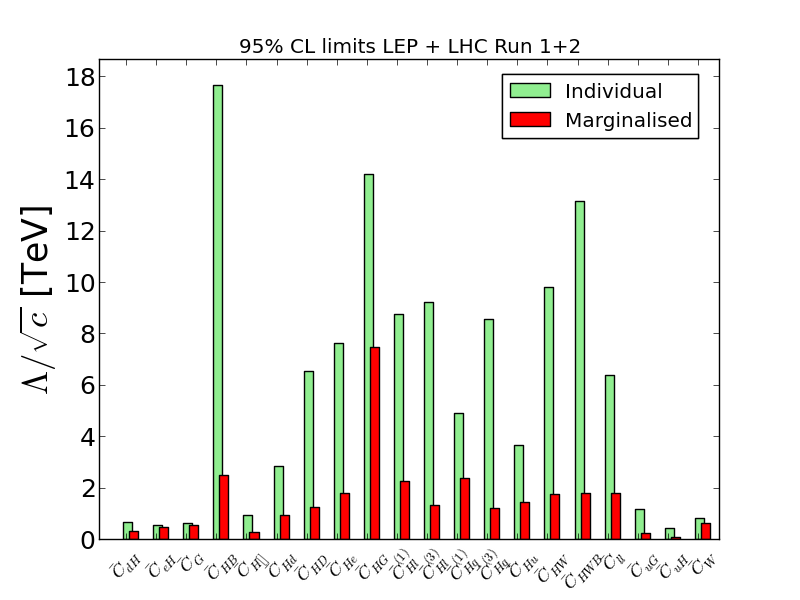}
 \caption{\it Summary of the 95\% CL bounds on the sensitivity (in TeV) %JE
 for an $\mathcal{O}(1)$ Wilson coefficient, obtained from marginalised (red) and individual (green) fits to the 20 dimension-6 operators entering in electroweak precision tests, diboson and Higgs measurements at LEP, SLC, and LHC Run 1 and 2.}
   \label{fig:barsummary}
\end{figure} 

The sensitivities to the scales of the operators in the Warsaw basis for an $\mathcal{O}(1)$ Wilson coefficient is summarised in Fig.~\ref{fig:barsummary}. %JE
Overviews of our results are shown in Figs.~\ref{fig:comparison2} (Warsaw basis)
and \ref{fig:comparison} (SILH basis), where we see in the upper panels
the results of fits where all the 20 operator coefficients are allowed to vary simultaneously, and in the
lower panels results where the operators are switched on one at a time. Fig.~\ref{fig:comparison2}
also shows comparisons with fits omitting the LHC Run 2 data, and Fig.~\ref{fig:comparison3}
displays explicitly the reductions in the uncertainties in the operator coefficients in the Warsaw basis when
the LHC Run 2 data are included.
Fig.~\ref{fig:comparison} (SILH basis) shows
a comparison with a fit to the precision $Z$-pole and $W$-mass data alone, and
a comparison with the results of~\cite{Ellis:2014jta}, which included Higgs results from Run 1
of the LHC only. Numerical results from the fits in the Warsaw and SILH bases are tabulated in
Table~\ref{tab:global}, and impacts of the different datasets on the global fit in the Warsaw basis
are shown numerically in Table~\ref{tab:impact}. Whereas the constraints from the precision electroweak observables have evolved
slowly, those from Higgs production are now much stronger than from Run 1, due to the
availability of much kinematical information as well as the increased statistics.
Correlations between the operator coefficients in the two operator bases are shown in Fig.~\ref{fig:correlations}.

Table~\ref{tab:fits} compares the qualities of the fits within the SM and the SMEFT,
displaying their respective $\chi^2$, $\chi^2 / n_{\text{d}}$, and $p$-values. The top line is for a fit to the SM and
the middle line is for a fit to the SMEFT allowing all 20 coefficients to vary, 
whilst the bottom line assumes a UV-completion of the SMEFT (indicated with with a $\star$)
that is renormalizable and weakly-coupled. These assumptions allow 13 coefficients to be non-zero, and in the Warsaw basis
the coefficients set to zero in this case are $C_G$, $C_W$, $C_{HG}$, $C_{HW}$, $C_{HB}$, $C_{HWB}$, and $C_{uG}$.
We see that neither the full SMEFT nor the SMEFT$^{\star}$ give fits that are significant improvements on the SM fit,
which has  already a very acceptable $p$-value. Thus, these fits provide no sign or evidence of any physics beyond the Standard Model.

%%%%%%
\begin{table}
\centering
 \begin{tabular}{| c | c | c | c |}
 \hline 
Theory & $\chi^2$ & $\chi^2 / n_{\text{d}}$ & $p$-value   \\ \hline \hline
SM & 157 & 0.987 & 0.532 \\ \hline
SMEFT & 137 & 0.987 & 0.528 \\ \hline
SMEFT$^{\star}$ & 143 & 0.977 & 0.564 \\ \hline
 \end{tabular}
  \caption{\it The $\chi^2$, $\chi^2 / n_{\text{d}}$, and $p$-values for the SM and SMEFT fits. 
  The middle line is a fit to the SMEFT allowing all 20 coefficients to vary, whilst the bottom line assumes a 
  UV-completion of the SMEFT that is renormalizable and weakly-coupled, indicated with a $\star$. 
  These assumptions allow just 13 non-zero coefficients, 
  and in the Warsaw basis the coefficients set to zero in this case are
  $C_G$, $C_W$, $C_{HG}$, $C_{HW}$, $C_{HB}$, $C_{HWB}$, and $C_{uG}$.}
  \label{tab:fits}
\end{table}
%%%%%%

Our new constraints on the dimension-6 operator coefficients can be applied to
variety of specific BSM scenarios. Specifically, we have studied extensions of the SM
that can make tree-level contributions to the operator coefficients, as tabulated in Table~\ref{tab:modeldefs},
using the dictionary proposed in~\cite{deBlas:2017xtg}: 
see Fig.~\ref{fig:model2param} and the numerical results in Section~\ref{sec:BSM}.
We have also explored the constraints imposed by the global fit on stops in the MSSM, which
contribute to the operator coefficients only at the loop level, see Fig.~\ref{fig:stops}.
These constraints are model-independent, and competitive with the model-dependent constraints
on stops from Run 2 of the LHC.

We can expect in the near future further substantial increases in the amounts of
information from diboson and Higgs production at the LHC as the ATLAS and CMS Collaborations
complete their analyses of data from Run 2. We emphasize the importance to SMEFT
analyses of making available as much information as possible on the kinematics of diboson
and Higgs production, since the $p_T$ and invariant mass distributions, in particular, are
more sensitive to dimension-6 operator coefficients than are the integrated production rates.
In this way maximal information can be extracted from the data and used, via a SMEFT
analysis, to constrain possible BSM scenarios, as we have illustrated in this paper. We cannot
know whether such an analysis will reveal any BSM physics, but in this way we will give the search
for new physics our best shot.

%%%%%%%%%%%%%%%%%%%%%%
\section*{Acknowledgements}

We thank Ilaria Brivio, Pier Paolo Giardino, Mart\'{i}n Gonz\'{a}lez-Alonso and Michael Trott for useful discussions, and Jérémie Quevillon for the MSSM Higgs mass calculations.
The work of JE was supported partly by the United Kingdom STFC Grant ST/P000258/1 and
partly by the Estonian Research Council via a Mobilitas Pluss grant.
VS acknowledges support from the Science and Technology Facilities Council (ST/P000819/1).
The work of CWM was supported by the United States Department of Energy under Grant Contract DE-SC0012704.
The work of TY was supported by a Junior Research Fellowship from Gonville and Caius College and partially supported by STFC consolidated grant ST/P000681/1.

%%%%%%%%%%%%%%%%%%%%%%
\appendix
\section*{Supplementary Material}
The complete $\chi^2$ function in both the SILH and Warsaw bases as well as all of our predictions made using $\mathtt{SMEFTsim}$ are available online at \url{https://quark.phy.bnl.gov/Digital_Data_Archive/SMEFT_GlobalFit/}.

%{\color{blue}The experimental measurements used in this fit constrain not just higher-dimensional operators, but light BSM physics as well. 
%For instance, consider a scalar singlet that may act as a mediator to a hidden sector through the Higgs portal, $\Delta \mathcal{L} = - \kappa_{\mathcal{S}} \mathcal{S} H^{\dagger} H$.
%There is a mixing angle, $\theta$, between the two mass eigenstates.
%In addition, the Higgs boson can decay to this new scalar if its mass is less than $m_h / 2$.
%In this model there is a universal modification of the Higgs signal strength (for decays to SM particles)
%\begin{equation}
%\frac{\sigma(i \to h)\, \text{Br}(h \to f)}{\sigma(i \to h)_{SM}\, \text{Br}(h \to f)_{SM}} = \cos^2\theta \left[1 - \text{Br}(h \to BSM)\right] \equiv x ;\quad i, f \in SM.
%\end{equation}
%The parameter $x$ is bounded by our fit.
%We find $x > \{0.98, 0.94, 0.90\}$ at the 1-, 2-, and 3-$\sigma$ levels, respectively.}

\bibliographystyle{utphys}
\bibliography{EMSY_v3}

\end{document}